\documentclass[twocolumn,times,final]{elsarticle}
\usepackage[latin9]{inputenc}
\usepackage{color}
\usepackage{array}
\usepackage{float}
\usepackage{amssymb}
\usepackage{graphicx}
\usepackage[unicode=true,
 bookmarks=false,
 breaklinks=false,pdfborder={0 0 1},backref=section,colorlinks=true]
 {hyperref}

\makeatletter

\providecommand{\tabularnewline}{\\}
\floatstyle{ruled}
\newfloat{algorithm}{tbp}{loa}
\providecommand{\algorithmname}{Algorithm}
\floatname{algorithm}{\protect\algorithmname}

\usepackage{cag}
\usepackage{framed}
\usepackage{multirow}
\usepackage{latexsym}

\usepackage{url}
\usepackage{xcolor}
\definecolor{newcolor}{rgb}{.8,.349,.1}

\journal{Graphics and Visual Computing}
\jnlurl{www.elsevier.com/locate/gvc}

\@ifundefined{showcaptionsetup}{}{%
 \PassOptionsToPackage{caption=false}{subfig}}
\usepackage{subfig}
\makeatother

\usepackage{listings}

\begin{document}

\begin{frontmatter}{}

\title{Volumetric Procedural Models for Shape Representation}

\author[1]{Andrew R. \snm{Willis}}

\author[2]{Prashant \snm{Ganesh}\fnref{fn1}\corref{cor1}}

\cortext[cor1]{Corresponding author: Tel.: +1-850-833-9350;}

\emailauthor{prashant.ganesh@ufl.edu}{Prashant Ganesh} 

\author[2]{Kyle \snm{Volle}\fnref{fn1}}

\author[1]{Jincheng \snm{Zhang}}

\author[3]{Kevin \snm{Brink}}

\fntext[fn1]{Equal Contributor}

\address[1]{University of North Carolina at Charlotte, Charlotte, 28223, USA}

\address[2]{University of Florida, 1350 N Poquito Road, Shalimar, 32579, USA }

\address[3]{Air Force Research Laboratory, Eglin Air Force Base, 32579, USA }
\begin{abstract}
This article describes a volumetric approach for procedural shape
modeling and a new Procedural Shape Modeling Language (PSML) that
facilitates the specification of these models. PSML provides programmers
the ability to describe shapes in terms of their 3D elements where
each element may be a semantic group of 3D objects, e.g., a brick
wall, or an indivisible object, e.g., an individual brick. Modeling
shapes in this manner facilitates the creation of models that more
closely approximate the organization and structure of their real-world
counterparts. As such, users may query these models for volumetric
information such as the number, position, orientation and volume of
3D elements which cannot be provided using surface based model-building
techniques. PSML also provides a number of new language-specific capabilities
that allow for a rich variety of context-sensitive behaviors and post-processing
functions. These capabilities include an object-oriented approach
for model design, methods for querying the model for component-based
information and the ability to access model elements and components
to perform Boolean operations on the model parts. PSML is open-source
and includes freely available tutorial videos, demonstration code
and an integrated development environment to support writing PSML
programs.
\end{abstract}

\end{frontmatter}{}


\section{Introduction}

The generation of realistic shape models for use in computer graphics
applications such as virtual reality, computer games and movie production
can be a time consuming task. These models often are built in a manner
similar to movie sets, i.e., facades are constructed which incorporate
the geometric structure and appearance necessary to provide a visually
convincing experience when photographed (movies) or experienced from
a limited viewpoint (games). However, important contexts exist where
the entire structure of the object(s) being modeled are of interest.
Examples of these contexts include most real-world simulation applications
such as physics modeling, finite element analysis, heat transfer analysis
and other contexts such as archaeological research or, more generally,
Building Information Modeling (BIM). This article details a new Procedural
Shape Modeling Language (PSML) which is a programming language for
specifying volumetric procedural models. The article describes PSML
and several contributions associated with the existence of a complete
language for procedural model specification in contrast to current
approaches which extend existing languages. A PSML interpreter which
executes PSML programs to generate 3D models is also described.

The initial goal of PSML is to provide programmers the ability to
describe objects as a semantic hierarchy of 3D shape elements where
each element may be a semantic group of objects, e.g., a floor of
a building, or an indivisible object, i.e., a brick within a building.
Each indivisible object, e.g., a brick from a building, is modeled
in terms of its geometry and appearance. Semantic groupings of objects,
e.g., a floor of a building are represented as a geometric 3D primitive
which serves to define the approximate scope, i.e., location, pose
and size, of elements that are considered to be part of the semantic
group. This hierarchical approach allows the programmer to make changes
at one level that propagate to other levels. For instance, changing
the dimensions of a brick can automatically adjust the element count
of a wall that uses that brick definition.
\begin{figure*}[t]
\begin{centering}
\subfloat[]{\begin{centering}
\includegraphics[height=1in]{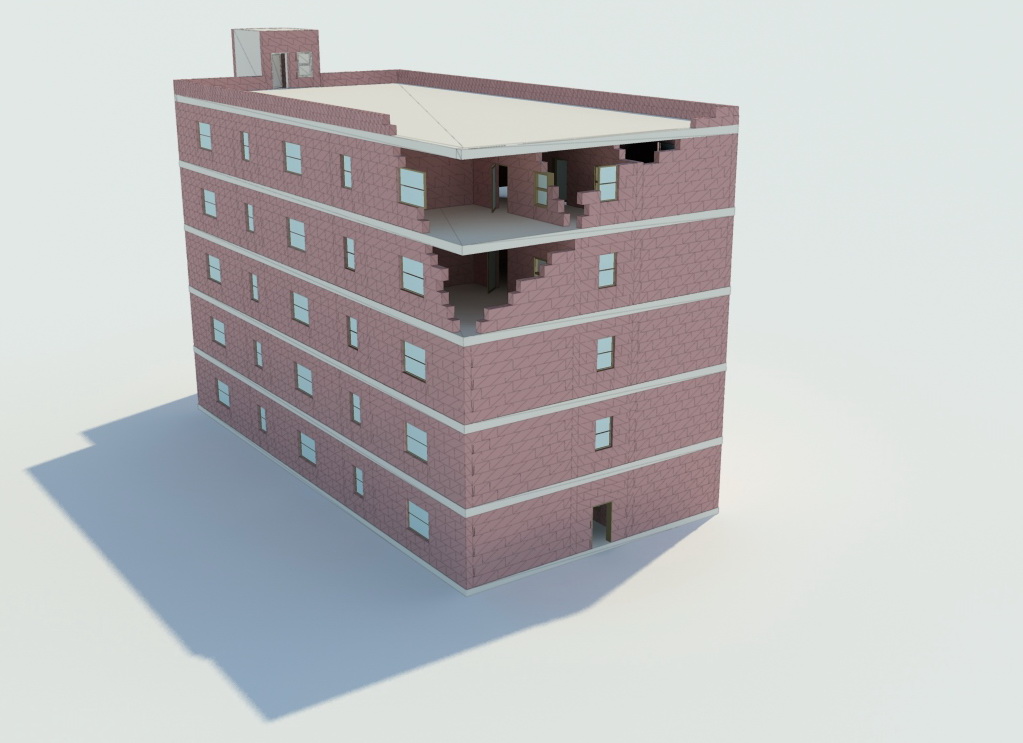}
\par\end{centering}
}\hfill{}\subfloat[]{\begin{centering}
\includegraphics[height=1in]{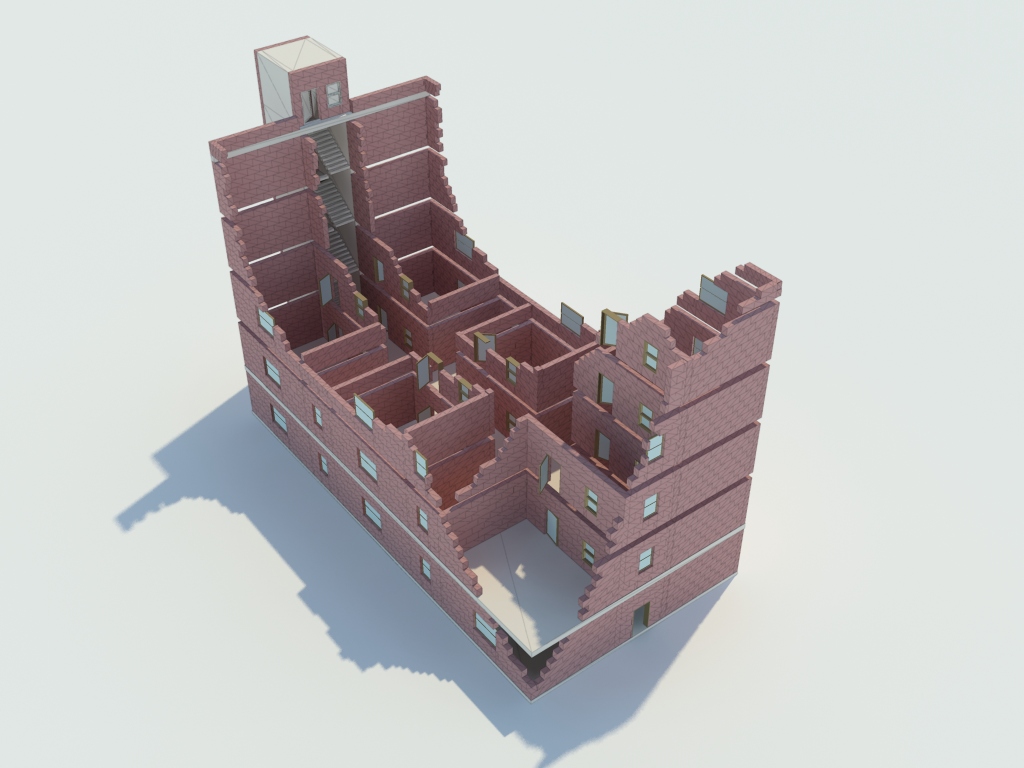}
\par\end{centering}
}\hfill{}\subfloat[]{\begin{centering}
\includegraphics[height=1in]{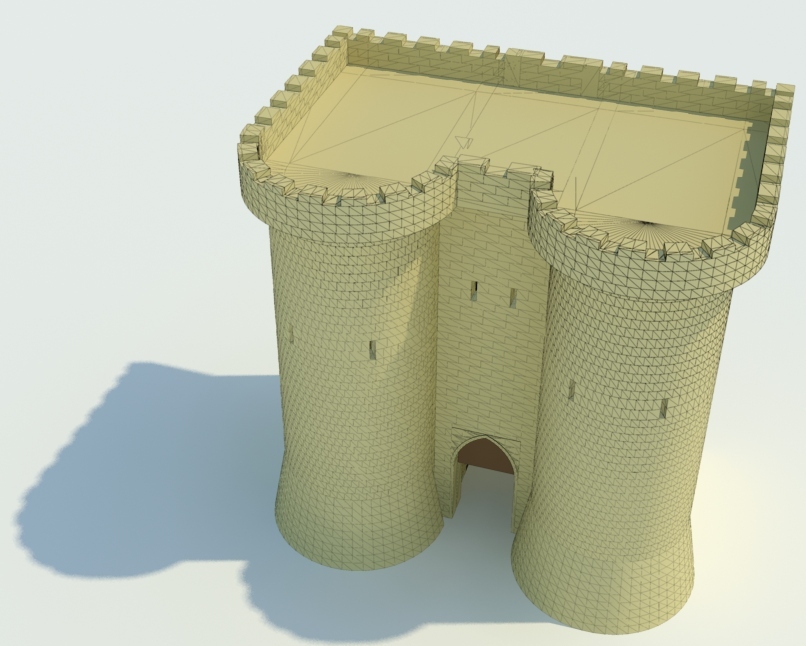}
\par\end{centering}
}\hfill{}\subfloat[]{\begin{centering}
\includegraphics[height=1in]{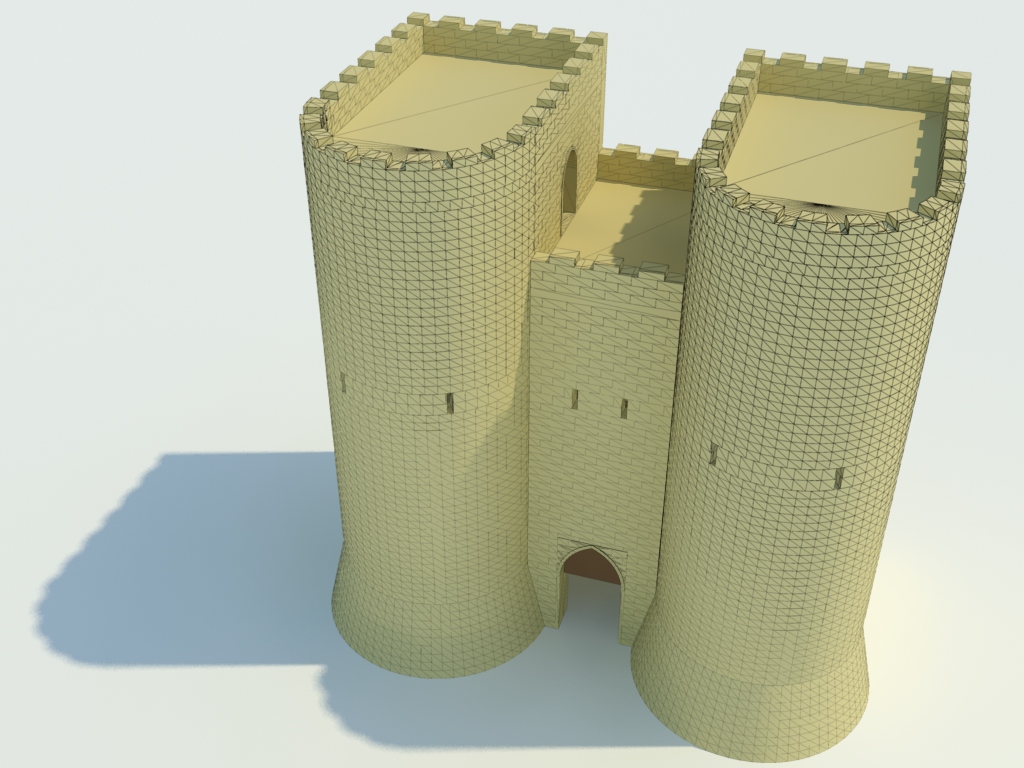}
\par\end{centering}
}
\par\end{centering}
\caption{\label{fig:TeaserFig}This article describes a volume-only based Procedural
Shape Modeling Language referred to as PSML which enables users to
generate volumetric 3D models. An office building (a,b) and a castle
gate (c,d) are shown which have been modeled exclusively in terms
of their volumetric elements using PSML.}
\end{figure*}

Modeling shapes in this way facilitates the creation of models that
more closely approximate the organization and structure of their real-world
counterparts. As such, users may query these models for information
defined over the semantic labels which is difficult to provide using
existing model-building approaches. For example, a building model
may include a large number of bricks and some volume of mortar needed
to bind these bricks together. PSML constructed models may be generated
which can readily extract quantities for this information in terms
of both the number of bricks and volume of mortar (as well as the
location and pose for all the \emph{brick }and \emph{mortar} elements).

Work on this topic was initially inspired from collaborative work
with anthropologists and archaeologists who require highly accurate
models for historically-important structures (typically architecture).
In this context, 3D models have the potential to be extremely useful
for addressing tasks which are typically both time-consuming and tedious
to perform in their current implementations. The most important of
these tasks is documenting and tracking the architectural elements,
i.e., the building blocks and structural adornments, of ancient structures.
In such contexts, cultural heritage researchers require 3D models
similar to those provided for Building Information Modeling (BIM),
i.e., the geometry of each indivisible element must be modeled.

This work describes a system that facilitates efficient generation
of models which can incorporate this information. As part of our approach,
we define a new language, PSML, which is similar in syntax and structure
to Java but also incorporates shape grammar programs as elements within
the sequential programming code. Considering that the language borrows
programming concepts with Java, users familiar with Java or other
object orientated programming will be able to easily learn and create
new objects. As in other procedural modeling implementations \citep{Wonka:InstantBuilding:2003,Mueller:ProceduralBuild:2006},
these grammars serve to replace shapes with other shapes using \emph{split()}
and \emph{repeat()} operations. The resulting models may be applied
for BIM systems as contemporary construction typically incorporates
structural regularity, i.e., indivisible objects that are similar
in geometry and appearance and whose relative positions and orientations
are often governed by a small set of simple rules. It is envisioned
that PSML could eventually be used for the documentation of historic
structures. However, the irregularities in both the indivisible elements
(since they are typically hand-made) and irregularities in the construction
commonplace to ancient architecture makes modeling such structures
difficult using the current implementation of PSML. To summarize,
the contributions of this article are as follows:
\begin{itemize}
\item We model shape as a hierarchical collection of 3D structures with
the intention of modeling objects from their large-scale sub-structures
down to their indivisible volumetric elements. As a result, generated
models are similar to their real-world counterparts and, as such,
may be interrogated to provide estimates of important quantities for
real-world objects.
\item The volumetric shape grammar associates labels to both object-space,
i.e. geometric objects, and void-space, i.e. the empty space that
bounds objects. These labels may be used to facilitate analysis that
requires knowledge of both types of information, e.g., furniture placement,
accessibility, navigation of virtual spaces, path-planning, etc.
\item We introduce a language that integrates sequential programming with
non-sequential and highly recursive shape grammar programming. Shapes
created in shape grammar programs may be referenced from and operated
on (using Boolean operations) in sequential code and shape grammars
may be initiated using shapes and variables defined from sequential
code. This integration enables new context-sensitive programming capabilities
which are otherwise difficult to address.
\item PSML supports an object-oriented approach for model design. Complex
models can be created by assembling simpler parts that are written
independently. This feature facilitates collaboration between designers
by allowing them to simultaneously work on independent parts. In addition,
PSML provides methods to select volumes defined in one grammar and
populate them with instances of other grammars, effectively ``injecting''
new designs into existing models without altering their code or introducing
dependencies to them.
\end{itemize}
Fig. \ref{fig:TeaserFig} serves to visually convey the central concept
of PSML and depicts some of instances of these capabilities that are
explained in more detail in the following sections. In (a,b) some
volume elements have been deleted to demonstrate that PSML allows
users to model structures at the building-block level and also captures
the relationships between external and internal structures such as
hallways, offices and stairs. (c,d) show two models from a single
castle gate shape grammar specified in PSML. These figures demonstrate
how high-level semantic relationships between shapes can be detected
and used to automatically transform the geometry to suit the circumstance.
(a) shows a castle gate with a high connected roof. Note the geometry
of the roof and battlements automatically adjust to splice together
smoothly. (b) shows a castle gate with separated roofs and a recessed
walkway in-between. Here, the geometry has been modified to suit the
lower walkway and doorways have been automatically added to the walkway
to allow for foot traffic (see \S~\ref{subsec:Void-space} for details).
Detecting and acting on the volumetric relationships between shapes
in this way is not possible using conventional, i.e., surface-based,
procedural models. 
\begin{figure}[h]
\centering{}\subfloat[]{\begin{centering}
\includegraphics[width=0.48\columnwidth]{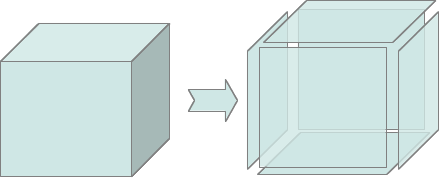}
\par\end{centering}
}\subfloat[]{\begin{centering}
\includegraphics[width=0.48\columnwidth]{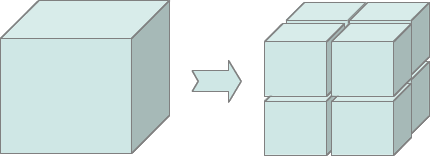}
\par\end{centering}
}\caption{\emph{\label{fig:ComponentSplit_and_VolumeSplit}}(a) Depicts the
component split operation on a cube which operates on the cube shape
as input and generates 6 faces of the cube as output. (b) Depicts
a split of a cube in PSML. PSML restricts shape grammar operations
such that all elements must be volumetric.}
\end{figure}

\section{Related Work}

Procedural modeling is becoming more popular and has been applied
in various fields, e.g., Plant modeling using L-Systems \citep{PRUSINKIEWICZ91LSys},
terrain modeling using fractals \citep{Belhadj:FractalTerrain:2005,Belhadj:FractalTerrain:2007},
model synthesis for generating large variations of given input models
\citep{Merrell:Synthesis:2007,Merrell:Synthesis:2008} and procedural
creation of Gothic window tracery using Generative Modeling Language
\citep{Havemann05PhdThesis,havemann_generativeGothic_2004}. Production
systems were generally used for generating highly detailed models
for large urban scenes which include cities \citep{Parish01ProceduralCity,CityEngine,groger2012citygml},
road networks within cities \citep{Chen:StreetModel:2008,Lipp:CityLayouts:2011},
the buildings within these cities \citep{Mueller:ProceduralBuild:2006,pottmann:freeform:2008}
and to generate graphics for games \citep{konecny2016procedural}.
Smelik in \citep{smelik2014survey} does a survey on the use of procedural
modeling for virtual worlds. The net effect of these tools and extensions
thereof, e.g., extruded surfaces as implemented in \citep{Kelly:ProcedrualExtru:2011},
coherence-based facade editing in \citep{Musialski:CoherenceFacade:2012}\emph{,
}automatic computation of the layout and organization of residential
buildings \citep{Merrell:FurnitureOpti:2010} and automatic placement
of furniture within the model as implemented in \citep{Merrell:InteractiveFurniture:2011,Germer:ProceduralFurniture:2009,Yu:FurnitureArrage:2011},
have demonstrated impressive capabilities for efficient generation
of geometric shapes.

Current implementations for grammar-based procedural models start
with geometric \emph{primitives} which are solid geometries that typically
come from a predefined set. Initial rules of the procedure create
instances of these primitives and place them into a relative configuration
to generate a \emph{mass model} (other approaches generate the mass
model by extruding a closed 2D polygon). The surfaces of the mass
model are then extracted from the mass model by decomposing each primitive
into its constituent planar faces using a \emph{component split} operation
(see Fig. \ref{fig:ComponentSplit_and_VolumeSplit}). Subsequent rules
then operate on the faces of the mass model to generate a 3D ``shell''
model. Such methods have been shown to produce highly detailed models
for building exteriors and, for appropriate definitions of the mass
model, can be used to model shells of both the interior and exterior
of a building, e.g., some of the Roman buildings of the virtual Pompeii
in \citep{Mueller:ProceduralBuild:2006} have courtyards. As such,
this work solves problems of great importance for generating massive
models of cities with highly detailed buildings. Other applications
of procedural modeling used in archaeology include \citep{kitsakis2017procedural,saldana2015integrated,dylla2008rome}.

However, ``shell'' models are not appropriate for more advanced
applications such as physical simulations and spatial analysis. Several
approaches concentrated on using volumetric grammars for such areas.
Whiting et al. created a volumetric grammar derived from CityEngine's
Computer-Generated Architecture (CGA) language \citep{Whiting:FeasibleBuilding:2009}.
This volumetric grammar is optimized to produce structurally sound
buildings. Terminals generated by this grammar are restricted to visible
volumetric shapes and no labeling of empty space is provided as it
is not required for structural optimization. Another work by Cutler
et al. addressed the creation of solid models \citep{cutler_procedural_2002}.
Cutler et al. introduced a procedural approach for converting closed
surfaces into solid models. The goal of their work, however, is restricted
to converting surfaces into volumes and does not aim to generate complicated
structured shapes and hierarchies. Similarly, Jesus et al. in \citep{jesus2016layered}
uses a layer based modeling technique which gives a structured approach
for an algorithm to merge two large grammars that are developed independently.
This allows for simpler syntax for larger models and writing grammar
with less lines of code. More recent work \citep{kutzner2020citygml}
has provided the ability to model solid and empty spaces in a hierarchical
framework but without some of the flexibility afforded by PSML.

The key difference between PSML and previous implementations is that
PSML is constrained to work \emph{entirely }from closed geometries
\emph{and} associates semantic labels to non-terminals, visible terminals
and empty space. In contrast to \citep{Whiting:FeasibleBuilding:2009},
PSML's ability to label and represent empty space volumes facilitates
the use of space information in both the creation process and as data
readily available for a post-processing step. The closed geometry
of terminal symbols generated by PSML can be converted into solid
geometry for further processing if needed. Thus, we view work of \citep{cutler_procedural_2002}
as a possible post processing step that may be applied to terminal
symbols. 

PSML's ability to create volumetric hierarchies with associated semantic
labels allows one to interrogate the model to ask important questions
such as the number and volume of some semantic element such as a brick.
This information allows the impressive models of ancient cities such
as ancient Pompeii \citep{Mueller:ProceduralBuild:2006} and other
ancient structures such as the Puuc building shown in \citep{Muller:PuucBuilding:2006}
to be interrogated for information that anthropologists and archaeologists
find useful. It is clear that these numbers will be approximate, yet,
development of high-fidelity models validated by cultural heritage
researchers shows promise for providing new insights on research on
these structures such as building methods, units of measure and construction
techniques employed by ancient civilizations. 

Another important application of procedural modeling is computer vision.
There has been a considerable amount of recent work that investigates
the use of shape grammars for vision tasks with a large number of
articles being produced that focus on segmentation of architecture
\citep{Koutsourakis09SingleViewReconstruction,Dick04ModelArchiImage,kyriakaki20144d}
within images or segmentation of building facade images \citep{Hohmann09CityFit,Zhao10ParsingArchitectureFacade,Muller07FacadeModeling,Teboul11SegmentFacade,Teboul10SegmentFacade}
with impressive results. However, these techniques were limited to
2D grammars since the labeled primitives produced by the used grammars
were limited to be 2D faces. It is envisioned that PSML may be an
effective context within which these vision algorithms may be applied.
The problem statement here is typically one where the user seeks to
infer the content of an image using a shape grammar program as a model
for the image content. The grammar constrains the solution space to
plausible organizations of semantic elements by requiring solutions
to come from the language of the grammar. In practice, this is accomplished
via an optimization algorithm that searches for the shape grammar
variable values the best ``fit'' the grammar shapes to the image
data. Not all grammars are equally amenable to this type of optimization
and there is interesting research into how to make grammars as functional
as possible \citep{jiang2018selection}. Work in \citep{jones2020shapeassembly}
represents a recent example of using machine learning to generate
structural shape programs that seek to be geometrically consistent
with their real-world counterparts. Other approaches use Monte Carlo
methods \citep{ritchie2015controlling,talton2011metropolis} to achieve
similar procedural modeling goals. These works also investigate the
opportunity for interpolating between generated models \citep{lienhard2017design}.
All of these applications may benefit from PSML as a new representation
for their shapes. 
\begin{algorithm}
\begin{lstlisting}[basicstyle={\scriptsize\ttfamily},breaklines=true,numbers=left,tabsize=2]
public class CoffeeMug extends ShapeGrammar { 
	float t = 0.4, w_top = 4.5, w_bottom = 3, h = 8.5; 
	
	public CoffeeMug() { 
		rules { 
			axiom::I(conicfrustum, {w_top, w_bottom, h}) R(Math.PI/2,0,Math.atan(h/(w_top-w_bottom)))T((w_top+w_bottom)/2,0,0)I(cylinder,{w_bottom, t}){vessel, handle_c}; 
			vessel::split(y, {t, scope.h - t}){vesselBody, vesselTop};
			vesselTop::split(r, {scope.r - t, t}){space, vesselBody};
			vesselBody::appearance(diffuse,{1,0,0}){terminal};
			handle_c::split(theta,{Math.PI,Math.PI}){space, handle_v};
			handle_v::split(r, {scope.r - t, t}){space, handle};
			handle::appearance(diffuse,{0,1,0}){terminal};
			space::void(){terminal}; 
		}
	}

	public static void main(String[] args) { 
		rules {
			Axiom::{CoffeeMug()}; 
		}
	}
}
\end{lstlisting}

\caption{\label{alg:SimpleCoffeeMug.psm}The contents of PSML file: CoffeeMug.psm}
\end{algorithm}
\begin{figure}[t]
\centering{}\subfloat[]{\begin{centering}
\includegraphics[height=0.65in]{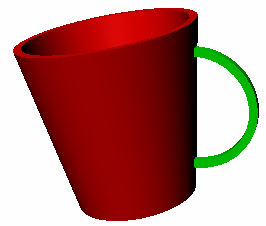}
\par\end{centering}
}\hfill{}\subfloat[t=1]{\begin{centering}
\includegraphics[height=0.65in]{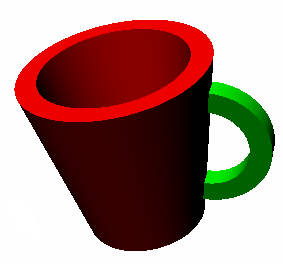}
\par\end{centering}
}\hfill{}\subfloat[w\_top=7]{\centering{}\includegraphics[height=0.65in]{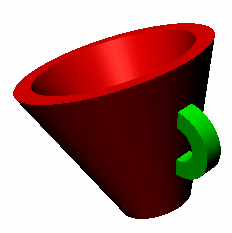}}\hfill{}\subfloat[h=10]{\begin{centering}
\includegraphics[height=0.65in]{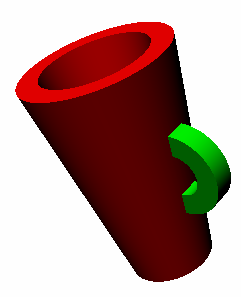}
\par\end{centering}
} \caption{\label{fig:PSML_CoffeeMugModels}(a) a visualization of the shape
generated by CoffeeMug.psm (Algorithm \ref{alg:SimpleCoffeeMug.psm})
(b-d) shows how varying PSML program variables introduce semantic
variations to the generated coffee mug models.}
\end{figure}

\subsection{Why not use an existing language directly?}

There exist languages for specifying sequential programs, e.g. C,
C++, Java, etc., and languages for specifying grammars, e.g. ANTLR,
YACC/Lex, Bison, etc. However, to our knowledge, there is no language
that incorporates both within a single source file. One benefit of
procedural models is their context-sensitive behavior. This refers
to instances when the generated geometry changes due to a specific
contextual condition, e.g., a visibility or occlusion constraint in
\citep{Mueller:ProceduralBuild:2006}. Detecting and acting on these
contexts requires sequential logic (scripting or special functions)
which serve to detect the context and change the geometry to suit
the specific situation. Hence, sequential logic is intrinsic to specifying
procedural models. On the other hand, procedural models are typically
specified as formal grammars and, as such, they require a syntax appropriate
for specifying formal grammars. Since the syntax for sequential code
and formal grammars are quite different a new language is proposed
which explores a specific combination of these two languages.

\subsection{Why not extend an existing language?\label{subsec:Why-not-extend}}

Published implementations for procedural modeling syntax have been
implemented as extensions to existing languages. For example, CityEngine
\citep{Parish01ProceduralCity,CityEngine} extends the Python scripting
language. In this context, users write python code and call shape
grammar (.cga) files which are associated with a separate interpreter
that executes the shape grammar and produces geometry. Such methods
have proven to be effective for generating shape as evidenced by the
popularity of this approach and the impressive results it has produced
\citep{Mueller:ProceduralBuild:2006}. Yet, from a programming perspective,
it can be difficult to develop within this context as the logic within
the Python code is separate from the logic applied for shape generation
in the CGA shape grammar. This separation between the Python script
and the shape grammar complicates the development of logic that bridges
this gap and requires the Python interpreter to communicate with the
shape grammar interpreter. The extent of compatibility between these
two interpreters can limit the possible operations between the sequential
logic and the shape grammar.

The fundamental reason that we do not extend an existing language
is based on the concept that PSML programs \emph{are }shapes, i.e.,
the fundamental object within the language is a shape. All PSML programs
are derived instances of this ``base class.'' In this regard, PSML
programs can be seen as a semantic collection of shape specifications.
The base class of an existing language, e.g., Java's \emph{Object
}class, is a core element upon which all abstractions are based. Redefinition
of such classes requires significant restructuring of the language
as a whole which motivated development of a separate complete language.
We are still investigating methods to extend existing sequential languages,
particularly Java, in ways that are tractable and still allow us to
maintain the desired abstractions and syntax presented as PSML in
this article.

\section{PSML Shape Grammars}

The approach PSML takes is to separate sequential statements from
the grammatical statements (rules) using simple syntax. PSML is a
combination of sequential code which has a structure and syntax inspired
by Java and shape grammar code which has a structure and syntax inspired
by L-systems \citep{PRUSINKIEWICZ91LSys}. The overall structure of
a PSML program includes one \emph{grammar }declaration that contains
one or more \emph{method }declarations. Each method declaration must
include at least one \emph{rules }declaration. A very basic example
of this structure is shown in Algorithm \ref{alg:SimpleCoffeeMug.psm}.

As mentioned in \S~\ref{subsec:Why-not-extend}, each grammar is
intended to represent some shape, i.e., the CoffeeMug.psm grammar
in Algorithm \ref{alg:SimpleCoffeeMug.psm} is intended to make geometric
models of coffee mugs as shown in Fig. \ref{fig:PSML_CoffeeMugModels}.
Typically, a collection of methods is defined within each grammar
and each of these methods takes as input a reference shape and an
optional list of arguments and operates on the reference shape, replacing
it with a sub-tree of other shapes generated from grammars defined
within that method. Arguments passed to methods allow each grammar
to use context-sensitive information to influence the shape generation
process which, among other things, is useful for controlling level-of-detail.
Rule blocks must be defined within each method and are used to specify
shape grammars. These grammars use the passed shape and argument variables
and locally defined variables to generate an instance of the grammar
shape, e.g., the grammar of Algorithm \ref{alg:SimpleCoffeeMug.psm}
generates a ``coffee mug.'' The algorithm defines a conic frustum
(line 6) with a hole (lines 7 and 8) that make up the body of the
cup. The handle is a green (line 12) cylinder (line 6) with a hole
removed from center (line 11) and cut in half (line 10), aligned with
and touching the side of the cup (as per its defined positioning).
Figure \ref{fig:PSML_CoffeeMugModels} shows various realization of
the coffee mug object and demonstrates how the representation enforced
shape constraints despite several parameter variations, e.g., the
handle and mug vessel surfaces connect consistently for these variations.
More generally, PSML's syntax for detecting the size and position
of the current volume allows the user to develop shapes that re-organize
their sub-components consistently over parametric variations, e.g.,
anisotropic scaling.

\subsection{PSML Programs}

Algorithm \ref{alg:SimpleCoffeeMug.psm} shows an example of a PSML
program file. As in Java, the program file name must match the name
of the grammar declaration which begins the program. The program body
is defined with braces '\{\}'. As in Java/C++/C, braces '\{\}' define
scope delimiters and are generically used to denote the beginning
and ending of execution blocks for flow control statements, e.g.,
if/else conditionals, for loops, while loops, etc. They also delimit
our newly defined \emph{rules} blocks which contain PSML shape grammar
rules. Scopes also serve to control variable resolution in a manner
similar to Java.

The body of each PSML grammar may include variable declarations and
initializations as well as a collection of methods which are analogous
to functions. Methods may only be called from shape grammar rules
and are invoked when they appear as a non-terminal successor symbol
within a rule of some shape grammar (see Rule Block and Rules for
details). The \emph{import} directive serves to import grammars and
their methods to another grammar and determines the collection of
methods which may be called from any grammar (similar to Java import
for classes). Rules which invoke grammar methods transfer control
to the statements of the selected method. 
\begin{figure*}[t]
\begin{centering}
\subfloat[]{\begin{centering}
\includegraphics[width=1.5cm,height=1.5cm,keepaspectratio]{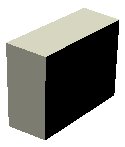}
\par\end{centering}
}\subfloat[]{\begin{centering}
\includegraphics[width=1.5cm,height=1.5cm,keepaspectratio]{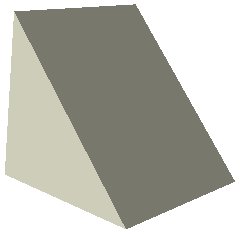}
\par\end{centering}
}\subfloat[]{\begin{centering}
\includegraphics[width=1.5cm,height=1.5cm,keepaspectratio]{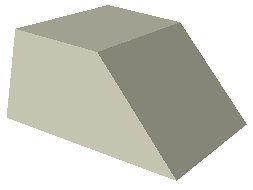}
\par\end{centering}
}\subfloat[]{\begin{centering}
\includegraphics[width=1.5cm,height=1.5cm,keepaspectratio]{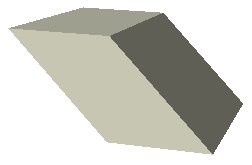}
\par\end{centering}
}\subfloat[]{\begin{centering}
\includegraphics[width=1.5cm,height=1.5cm,keepaspectratio]{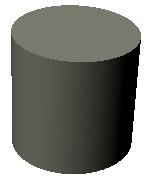}
\par\end{centering}
}\subfloat[]{\begin{centering}
\includegraphics[width=1.5cm,height=1.5cm,keepaspectratio]{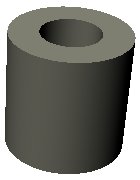}
\par\end{centering}
}\subfloat[]{\begin{centering}
\includegraphics[width=1.5cm,height=1.5cm,keepaspectratio]{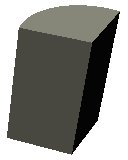}
\par\end{centering}
}\subfloat[]{\begin{centering}
\includegraphics[width=1.5cm,height=1.5cm,keepaspectratio]{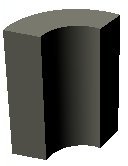}
\par\end{centering}
}\subfloat[]{\begin{centering}
\includegraphics[width=1.5cm,height=1.5cm,keepaspectratio]{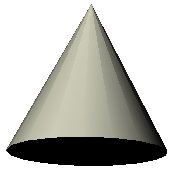}
\par\end{centering}
}
\par\end{centering}
\begin{centering}
\subfloat[]{\begin{centering}
\includegraphics[width=1.5cm,height=1.5cm,keepaspectratio]{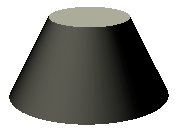}
\par\end{centering}
}\subfloat[]{\begin{centering}
\includegraphics[width=1.5cm,height=1.5cm,keepaspectratio]{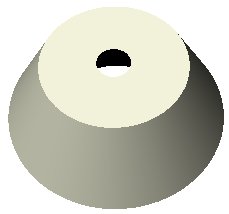}
\par\end{centering}
}\subfloat[]{\begin{centering}
\includegraphics[width=1.5cm,height=1.5cm,keepaspectratio]{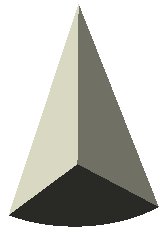}
\par\end{centering}
}\subfloat[]{\begin{centering}
\includegraphics[width=1.5cm,height=1.5cm,keepaspectratio]{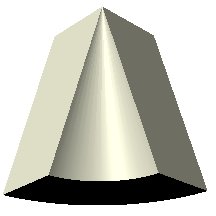}
\par\end{centering}
}\subfloat[]{\begin{centering}
\includegraphics[width=1.5cm,height=1.5cm,keepaspectratio]{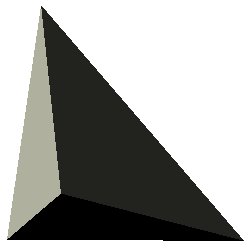}
\par\end{centering}
}\subfloat[]{\begin{centering}
\includegraphics[width=1.5cm,height=1.5cm,keepaspectratio]{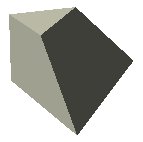}
\par\end{centering}
}\subfloat[]{\begin{centering}
\includegraphics[width=1.5cm,height=1.5cm,keepaspectratio]{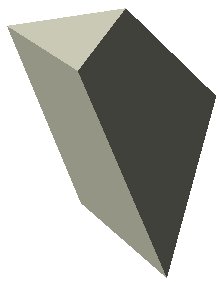}
\par\end{centering}
}\subfloat[]{\begin{centering}
\includegraphics[width=1.5cm,height=1.5cm,keepaspectratio]{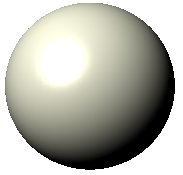}
\par\end{centering}
}\subfloat[]{\begin{centering}
\includegraphics[width=1.5cm,height=1.5cm,keepaspectratio]{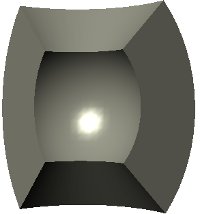}
\par\end{centering}
}\subfloat[]{\begin{centering}
\includegraphics[width=1.5cm,height=1.5cm,keepaspectratio]{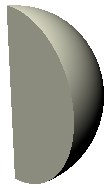}
\par\end{centering}
}
\par\end{centering}
\caption{\emph{\label{fig:PSML_PrimitiveShapes}}(a-s) depict 19 different
volumetric primitives which may be used in PSML programs to construct
3D shapes. These shapes are generated by splitting one of six basic
primitive shapes which are (a) the box, (b) the ramp, (e) the cylinder,
(i) the cone, (n) the tetrahedron, and (q) the sphere. Other shapes
are derived from these basic shapes as follows: (c,d) are derived
from (b), (f-h) are derived from (e), (j-m) are derived from (i),
(o-p) are derived from (n), and (r, s) is derived from (q).}
\end{figure*}

\subsection{Rule Blocks and Rules}

Each method must include a rules block which is denoted by the keyword
\emph{rules }and has a scope as defined by matching braces. Each rules
block contains a set of shape grammar production rules where each
production rule has the generic form as specified in (\ref{eq:syntax}). 

\begin{equation}
predecessor:condition:successor\,;\label{eq:syntax}
\end{equation}

Execution of a production rule causes the non-terminal symbol referred
to as the \emph{predecessor} to be replaced by the \emph{successor}
which may be one or more terminal or non-terminal symbols. The \emph{condition
}term is a branching expression: if \emph{condition }evaluates as
true, then the rule is executed; else, the next instance of a production
rule having the same predecessor is executed. The syntax of PSML rules
are inspired by those described for L-systems in \citep{PRUSINKIEWICZ91LSys}
but differ by not allowing for in-line scripts within rules. These
scripting functions have been replaced by the availability of successor
\emph{methods.}

Multiple rules may be written for a given non-terminal symbol to allow
for distinct behaviors based on the evaluation of the rule \emph{condition.}
Precedence for these rules is defined by the order that these rules
appear within the grammar, where the first rule for the symbol has
highest precedence. One may designate a new shape to be a real-world
object by indicating the shape is terminal using the special string
\emph{terminal} (see Algorithm \ref{alg:SimpleCoffeeMug.psm}, lines
8, 11, 12). Terminal symbols do not appear as predecessors in any
production rule and are visible elements that exist in the final 3D
model with the exception of terminals declared to be void() (see \S~\ref{subsec:Rule-Functions},
\emph{Modeling void-space} for details). While the treatment of non-terminal
symbols in PSML is similar to \citep{krecklau_generalizedNonTerminals_2010},
a key difference is PSML's ability to associate semantic labels to
terminal and non-terminal symbols which allows further analysis to
be conducted based on the semantic meaning of shape elements.

The way PSML processes rule blocks has several advantages. Since blocks
are part of the language itself, rules have direct access to variables
available within the sequential scope in which they are defined. In
addition, the rule blocks themselves can be regarded as variables
within the same scope, allowing subsequent sequential code to access
and post-process shapes created in an earlier rules block.

\subsection{\label{subsec:Rule-Functions}Rule Functions}

Each rule within a shape grammar may include a sequence of one or
more special functions which are referred to as rule functions. These
functions are inserted before the list of successor symbols and serve
to transform the predecessor shape into the successor shape(s), e.g.,
changing the shapes diffuse reflectance or color (see Algorithm \ref{alg:SimpleCoffeeMug.psm},
lines 8, 11). PSML rule functions include special functions for creating
and transforming (rotating/translating/scaling) objects as well as
the $split()$ and $repeat()$ operations as described in \citep{Wonka:InstantBuilding:2003}
and \citep{Mueller:ProceduralBuild:2006}. However, noticeably absent
from the list of available operations is the component split or \emph{$Comp()$}
operation which divides 3D volumes into shapes of lesser dimension,
i.e., faces, edges or vertices (see Fig. \ref{fig:ComponentSplit_and_VolumeSplit}(a)
for an example of a face split of a cube). Omission of this function
effectively constrains all our grammar shapes to remain volumetric
and preserves the semantic interpretation of each terminal as a virtual
object. This tends to produce more complex models, yet these models
encode important information which allows PSML to perform operations
that may be difficult to deal with using shell-based representations.
The following sections provide a complete list of available rule functions. 

\subsubsection{Creating Shapes Instances}

This function replaces the predecessor shape with a completely new
shape and has syntax: I(String \emph{type}, double{[}{]} \emph{params})
where \emph{type} indicates the shape to be created and \emph{params
}define the parameters necessary to create the shape. The function
may be invoked multiple times and successor symbols are associated
with each created shape in the same sequence as they are created within
the rule. For example, in Algorithm \ref{alg:SimpleCoffeeMug.psm}
(line 5) a conic frustum is created and associated to the symbol \emph{vessel
}and a cylinder is created and associated to the symbol \emph{handle\_c.}

All instances draw from a pre-defined set of 3D shape primitives.
There are six basic primitives: \{\emph{box, cylinder, sphere, cone,
ramp, tetrahedron}\}. Each primitive may be subdivided/split along
its principle axes into other derived shapes, making it possible to
create 19 different primitives as shown in Fig. \ref{fig:PSML_PrimitiveShapes}.
For example, a conic ring (the side of the coffee mug) is generated
on line 7 by splitting the conic frustum into two parts having different
radii and the mug handle is generated on line 10 by splitting a cylinder
into a ring (the handle) and a space. Triangulations of these primitive
shapes are pre-defined in the PSML built-in function library. 

Similar to \citep{edelsbrunner2017procedural}, PSML supports cylindrical,
and spherical coordinate systems in addition to the Euclidean coordinate
system. The Euclidean coordinate system is used for boxes, ramps,
and tetrahedrons. The cylindrical coordinate system is used for cylinders
and cones, and the spherical coordinate system is used for spheres
and enables PSML to model architecture with round geometry.

\subsubsection{Split/Repeat}

These functions replace the predecessor shape with a sequence of new
shapes and require two arguments. The first argument for these functions
indicates the coordinate axis of the the split, i.e., new shapes will
be defined by slicing through the predecessor shape with a plane perpendicular
to the indicated axis direction. The second argument determines where
these divisions will spatially occur along that axis (see Algorithm
\ref{alg:SimpleCoffeeMug.psm} lines 6, 7, 9, 10). The repeat function
takes in an optional third argument which is used to introduce an
offset for the initial division which is particularly useful in structures
with offsets such as sequential rows of bricks in a wall.
\begin{algorithm}
\begin{lstlisting}[basicstyle={\scriptsize\ttfamily},breaklines=true,numbers=left,tabsize=2]
public class Bricks extends ShapeGrammar { 
	float[] defaultBrickDims = {0.4, 0.25, 0.2, 0};
	public Bricks() {
		float isEuclidean=
				myShape instanceof Shape3D.CartesianShape;
		float isCylindrical=
				myShape instanceof Shape3D.RotaryShape;
		rules { 
			parent::repeat(y,{@brick.height,@brick.height},0){even, odd};
			even:isEuclidean:repeat(x, {@brick.width},0) {brick};
			odd:isEuclidean:repeat(x,{@brick.width},@brick.width/2){brick};
			even:isCylindrical:repeat(theta,{@brick.width/myShape.r},0){brick};
			odd:isCylindrical:repeat(theta,{@brick.width/myShape.r},(@brick.width / myShape.r)/2){brick};
			brick::appearance(texture,@brick.texture)
					appearance(specular, {0,0,0}){terminal};
		}
	}
	public static void main(String[] args) { 
		rules {
			axiom::T(5, 0, 0)I(box, {4, 8, 0.5})
					I(conicfrustum, {1.5, 2, 8})
					T(-5, 0, 0)I(cylinder, {2, 8})
					{wall, cone, cyl};
			cone::split(r, {1.5, 0.5}){space, wall};
			cyl::split(r, {1.5, 0.5}){space, wall};
			wall::useAttributes(brick.properties, sand){Bricks()};
			space::void(){terminal};
		}
	}
}
\end{lstlisting}

\caption{\label{alg:Bricks.psm}The contents of a PSML file: Bricks.psm. Discussion
of the program is provided as part of Fig. \ref{fig:Demo-of-useAttributes-and-myShape}.}
\end{algorithm}
\begin{algorithm}
\begin{lstlisting}[basicstyle={\scriptsize\ttfamily},breaklines=true,numbers=left,tabsize=2]
attributes sand {
	brick.width = 1.5;
	brick.height = 0.6;
	brick.texture = sandStone.jpg;
}
attributes rock { 
	brick.width = 0.7;
	brick.height = 0.3;
	brick.texture = rock.jpg;
}
\end{lstlisting}

\caption{\label{alg:Bricks.Attributes}The contents of PSML attribute file:
brick.properties referenced on line 26 of Bricks.psm from Algorithm
\ref{alg:Bricks.psm}.}
\end{algorithm}

Note that split operations act along axes of the coordinate system
defined for each shape. Rules may be adapted to different primitives
by specifying primitive-sensitive code, either using sequential flow-control
statements, e.g., if/then, or using conditional rules that only execute
on specific primitive types.
\begin{figure}
\begin{centering}
\hfill{}\subfloat[]{\begin{centering}
\includegraphics[height=0.7in]{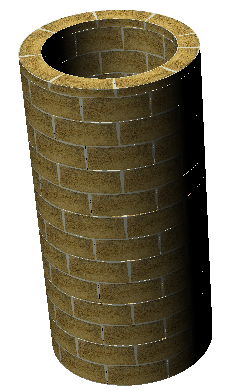}
\par\end{centering}
}\hfill{}\subfloat[]{\begin{centering}
\includegraphics[height=0.7in]{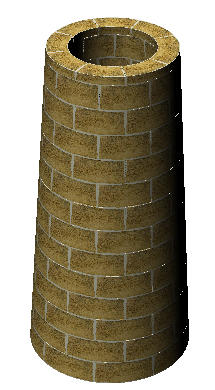}
\par\end{centering}
}\hfill{}\subfloat[]{\begin{centering}
\includegraphics[height=0.7in]{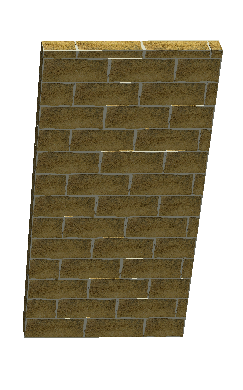}
\par\end{centering}
}\hfill{}\hfill{}\subfloat[]{\begin{centering}
\includegraphics[height=0.7in]{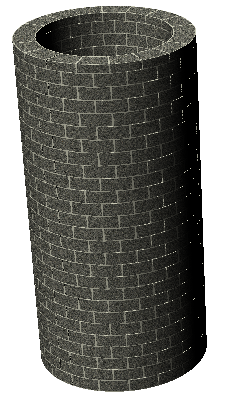}
\par\end{centering}
}\hfill{}\subfloat[]{\begin{centering}
\includegraphics[height=0.7in]{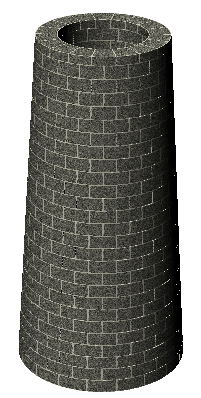}
\par\end{centering}
}\hfill{}\subfloat[]{\begin{centering}
\includegraphics[height=0.7in]{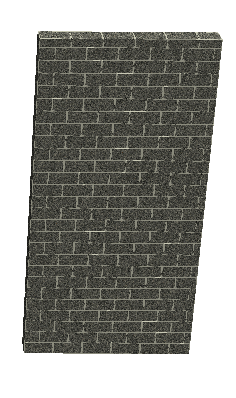}
\par\end{centering}
}\hfill{}
\par\end{centering}
\caption{\label{fig:Demo-of-useAttributes-and-myShape}(a) a sequence of models
generated by running the PSML program Bricks.psm (Algorithm \ref{alg:Bricks.psm}).
(b) demonstrates the utility of the \emph{useAttributes()} function,
in particular, this model is generated by substituting ``rock''
for ``sand'' on line 26. Lines 4, 5 show the utility of the \emph{instanceof
}operator for determining the coordinate system of the passed primitive
which enables distinct methods for placing the bricks that depends
on the coordinate system of the passed primitive. Lines 7-10 demonstrate
the utility of attributes and their syntax when used within a method.}
\end{figure}

\subsubsection{Spatial Transformations}

These functions operate on the predecessor shape by changing the linear
transformation that determines the size, position and orientation
of the object. There are three transformation operations: \emph{scaling,
rotation }and \emph{translation. }The scaling function is denoted
by S($s_{x},s_{y},s_{z}$) and scales the predecessor shape by adjusting
its bounding box in $xyz$ by a factor given by $s_{x}$, $s_{y}$
and $s_{z}$ respectively. The rotation function is denoted by R($\theta$,$\phi$,$\psi$)
and specifies a rotation matrix as a sequence of axis-aligned rotations
(Euler angles) around the $xyz$ axes; in that order. The translation
function is denoted T($t_{x},t_{y},t_{z}$) and translates the predecessor
shape by moving the center of its bounding box in $xyz$ by an amount
of $t_{x}$,$t_{y}$ and $t_{z}$ respectively. These functions may
be invoked in combination with instancing or constructor functions
to create and place primitives which is often a part of the axiom
rule of a PSML shape grammar.

\subsubsection{Appearance\label{subsec:Appearance}}

A generic\emph{ }function is provided to alter the appearance of the
predecessor shape. The function is denoted \emph{
\[
appearance(attrType,\,\,attrValue)
\]
 }where $attrType$ is a string value that specifies the type of appearance
attribute and $attrValue$ denotes the values that the attribute will
take. The following list describes the attribute types which may be
modified directly in PSML: \emph{color}, \emph{material}, \emph{ambient},
\emph{emissive}, \emph{diffuse}, \emph{specular}, \emph{shininess},
\emph{bump, bumpmap}, \emph{bumpweight}, \emph{texture}, \emph{transparency}.

\subsubsection{Void-space\label{subsec:Void-space}}

A special function \emph{$void()$} is used to make the associated
symbol an invisible terminal in the final model (see Algorithm \ref{alg:SimpleCoffeeMug.psm}
line 13). Voids specified in PSML models are particularly useful for
void-space analysis, i.e., detecting context-sensitive situations
that require special-purpose placement of semantic objects. As mentioned
earlier, this ability to label empty space terminals is one of the
key differences from previous volumetric approaches such as \citep{Whiting:FeasibleBuilding:2009,krecklau_generalizedNonTerminals_2010,Leblanc:ComponentModeling:2011}.
Formal definitions for these spaces and the ability to query and operate
on these spaces afforded by PSML shows promise for enabling new approaches
for automatic building layout design \citep{Merrell:FurnitureOpti:2010}
and automatic placement of furniture \citep{Merrell:InteractiveFurniture:2011,Germer:ProceduralFurniture:2009,Yu:FurnitureArrage:2011}.
Examples of uses for void spaces are shown in Fig. \ref{fig:TeaserFig}(c,
d), \ref{fig:Office-building} and \ref{fig:BooleanDemo.psm-results}
and will be further discussed in the Results section. 

\subsection{Attributes}

Procedural shape models often include a large number of constants
used to define object attributes such as their appearance, size, position,
adornment, etc. In many cases specific choices for these constants
determine the ``look-and-feel'' of the generated shapes. Since the
constants merely determine the starting values for variables within
the shape grammar, we created attribute files as a mechanism for making
these constants external to the shape program.

Inclusion of attribute files allows users to interact with the shape
generation process without the requirement that they understand PSML
code. In this context, users modify the attribute files and generate
instances of different shapes from the grammar. This provides inexperienced
users methods to control the shape generation process and give their
models visually distinct styles (see Fig. \ref{fig:Demo-of-useAttributes-and-myShape}).
We refer to these constants collectively as attributes and PSML provides
a rule function, \emph{
\[
useAttributes(filename,attrGroup)
\]
} where $filename$ references an external attribute file such as
that shown in Algorithm \ref{alg:Bricks.Attributes} and $attrGroup$
denotes a specific collection of attributes to load from the attribute
file. Each call to this function loads a collection of these constants
from a file and makes their values available to subsequent rules.
Context-sensitive uses of this rule function allow a single shape
grammar to exhibit different geometry and appearance depending upon
the context of the rule. After invoking useAttributes, the loaded
values are referenced using the prefix '@', and then writing the attribute
name (see Algorithm \ref{alg:Bricks.psm}: line 26 loads an attributes
file, lines 7-10 reference brick dimension attributes. Algorithm \ref{alg:Bricks.Attributes}
shows the contents of referenced attribute file).

\subsection{The myShape and scope variables and Namespaces}

As mentioned earlier, methods may be invoked from PSML shape grammar
rules. These methods are implicitly passed the shape of the appropriate
successor symbol for the invoking rule. This shape is assigned as
a special \emph{Shape} variable in the invoked method where the identifier
of the variable is \emph{myShape. }The \emph{myShape }variable is
similar to the \emph{this }reference available in many object-oriented
programming languages and carries with it many important attributes
of the shape passed to the method. These include the \emph{primitive
type} of the passed shape, the \emph{coordinate system }of the passed
shape and \emph{values} for all of the parameters which define the
shape. Namespace variables may be applied to detect the primitive
type of the passed primitive shape using the \emph{$instanceof$ }operator.
This operator is intended to behave in exactly the same way as the
Java \emph{instanceof} operator. However, this operator serves to
detect instances of different predecessor shapes or to detect the
preferred coordinate system of the predecessor shape. There are 22
namespace variables, 19 of these account for each primitive shape
type as shown in Fig. \ref{fig:PSML_PrimitiveShapes} and an additional
3 are provided which detect only the coordinate system type of the
passed primitive type (see Algorithm \ref{alg:Bricks.psm} lines 4,
5). Sequential statements may evaluate expressions that use the \emph{myShape
}variable to perform context-specific rule blocks as shown Fig. \ref{fig:Demo-of-useAttributes-and-myShape}.

The \emph{scope} variable is very similar to \emph{myShape. }The difference
is that \emph{scope} refers to the predecessor of a rule. This allows
access to the dimensions of the immediate predecessor in each rule
and facilitates the construction of recursive grammars as explained
in the next section.

\subsection{Recursion}

\begin{figure}
\begin{centering}
\subfloat[]{\begin{centering}
\includegraphics[viewport=0bp -1in 450bp 289bp,width=0.28\columnwidth]{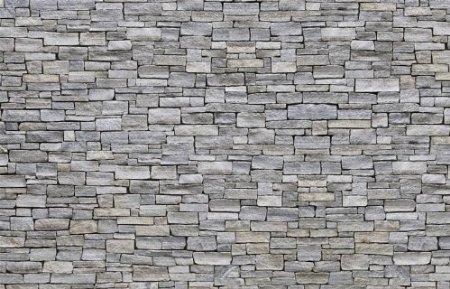}
\par\end{centering}
}\subfloat[]{\begin{centering}
\includegraphics[viewport=0bp 0bp 1024bp 1024bp,width=0.28\columnwidth]{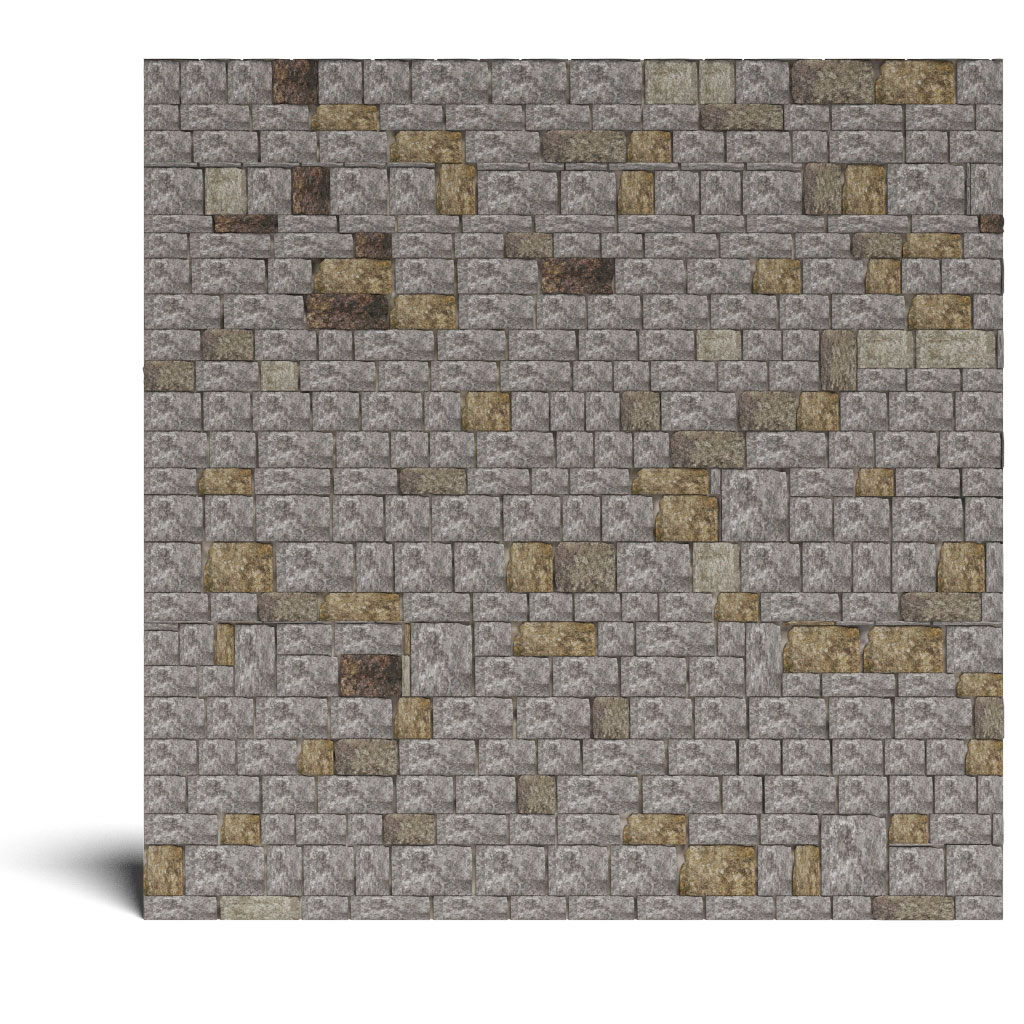}\includegraphics[viewport=0bp 0bp 1024bp 1024bp,width=0.28\columnwidth]{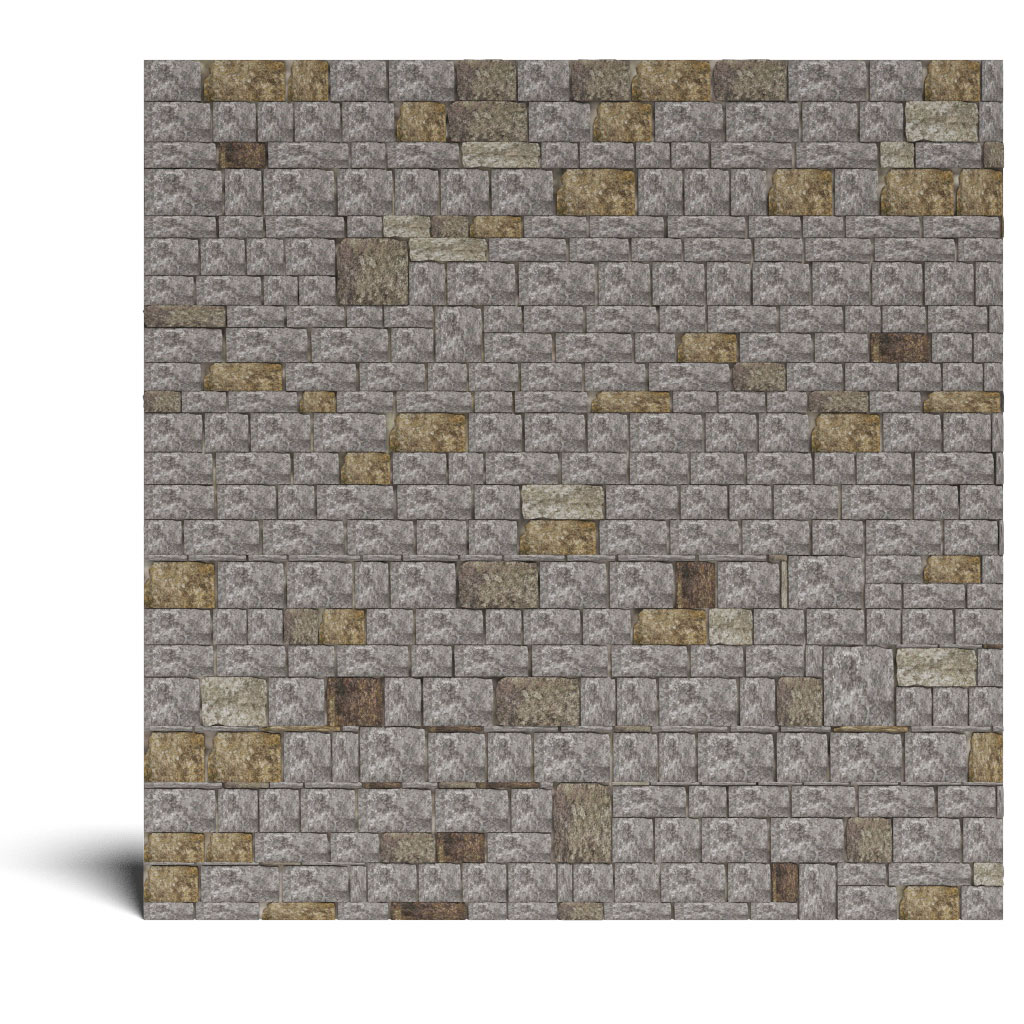}
\par\end{centering}
}
\par\end{centering}
\caption{\label{fig:recursion-irregular-patterns}Irregular patterns: a) is
an image of an actual stone wall in reality with irregular patterns.
b) shows two of our realizations of a randomized stone wall using
a randomized recursive grammar.}
\end{figure}

\begin{figure}
\begin{centering}
\subfloat[]{\begin{centering}
\includegraphics[angle=90,height=4cm]{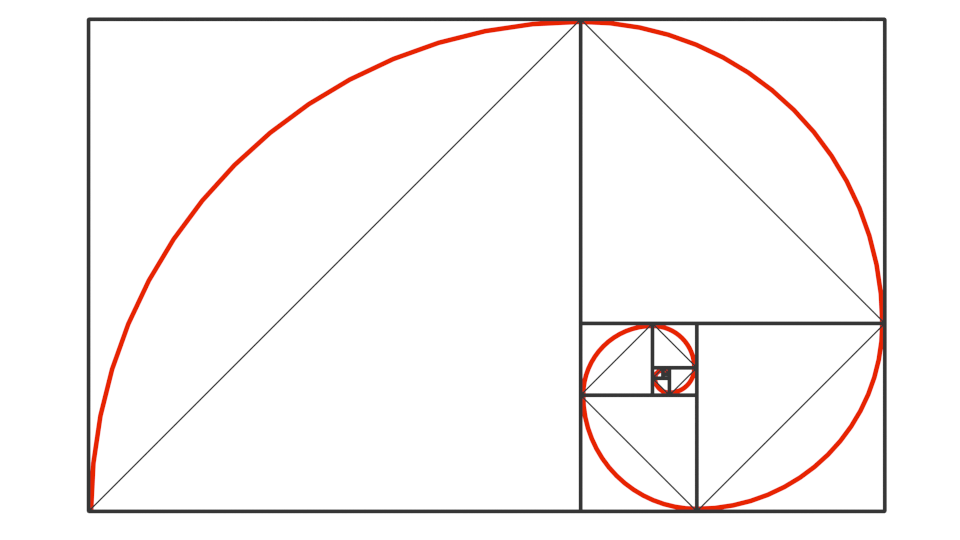}
\par\end{centering}
}\subfloat[]{\begin{centering}
\label{fig:RecursiveGrammar2}\includegraphics[height=4cm]{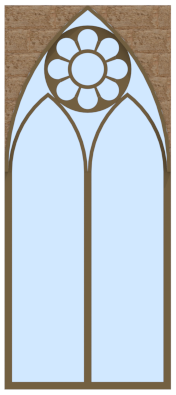}
\par\end{centering}
}\subfloat[]{\begin{centering}
\label{fig:RecursiveGrammar4}\includegraphics[height=4cm]{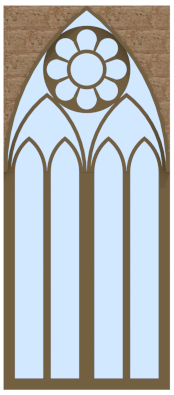}
\par\end{centering}
}
\par\end{centering}
\caption{\label{fig:Recursion-Gothic Window}Recursive nested patterns: (a)
a Fibonacci spiral generated by a 7-line recursive grammar. (b) a
Gothic window generated with two levels of recursion. (c) a Gothic
window generated with three levels of recursion.}
\end{figure}

Recursion is an intrinsic property of grammars. A recursive grammar
is one that has a non-terminal $\beta$ that produces further down
the derivation tree a successor that is $\beta$. Recursion allows
PSML to produce complicated shapes, such as irregular patterns (Fig.
\ref{fig:recursion-irregular-patterns}), or nested patterns (Fig.
\ref{fig:Recursion-Gothic Window}) that are hard to achieve using
sequential code such as \citep{Leblanc:ComponentModeling:2011}. PSML
provides the scope variable as means to check the attributes of the
current non-terminal and stop the recursion if some criteria is met,
which is used to terminate the shape derivation for \ref{fig:Recursion-Gothic Window}(a).

\subsection{Execution Model}

Statements are divided into sequential statements, i.e., Java-style
code, and rule blocks, i.e., shape grammar code. Execution begins
at the \emph{main()} method of the designated initial PSML grammar.
The \emph{main() }method is passed the unit cube as it's initial shape.
Statements in each method are executed sequentially until a shape
grammar is encountered as indicated by a \emph{rules }block. Within
each rules block, a shape grammar is defined which consists of a collection
of rules. The first rule of the grammar is taken as the \emph{axiom
}of the grammar, i.e., the initial production rule which is applied
using the shape passed to the method as a predecessor. Execution of
the shape grammar continues by looking up rules/methods to expand
each successor shape until all of the symbols are terminal symbols.
The specific sequence of executed production rules generates a \emph{parse
tree. }This tree records the sequence of production rules as a hierarchy
where non-terminal symbols are the internal nodes of the tree and
terminal symbols are the leaves of the tree. As in \citep{Mueller:ProceduralBuild:2006},
a scenegraph is used to track the sequence of grammatical productions
executed to generate each terminal object, i.e., the parse tree (or
derivation tree). The resulting scenegraph includes a hierarchy of
semantically meaningful groupings of objects from very large collections
of objects in the vicinity of the root, e.g., the floor of a building,
down to indivisible elements at the leaves, e.g., a brick of a building.
The end of each rule block may be followed by additional code that
may query the scenegraph using the \emph{terminals() }or \emph{instances()
}functions and operate on the returned symbols to make decisions regarding
the addition of model details or take other needed context-specific
actions (\S~\ref{sec:Working-with-Symbols} explains this in more
detail). Execution ends when the last statement of the \emph{main()
}method has been executed.

The execution model for the shape grammars is depth-first, i.e., the
interpreter expands the first successor completely to all of its terminal
elements before proceeding to the next symbol. This choice of execution
model for PSML limits some functionality afforded by priority-based
execution models as described in \citep{Mueller:ProceduralBuild:2006}.
Specifically, priority-based grammars allow the programmer to require
specific rules to be evaluated (executed) before other rules are evaluated.
This may be used to create geometry that will impact a context sensitive
criterion such as visibility tests in subsequent rules. However, these
limitations are somewhat offset by the fact that correct utilization
of priority values within priority-based grammars requires the user
to be aware of the possibly complex relationships between the priorities
assigned to all of the rules in order to predict the model construction
outcome. For models involving a large number of grammars, each of
which may incorporate a separate priority scheme, tracking the priority
of a given rule for a specific derivation can be complex for both
the programmer and interpreter. Hence extensive use of priorities
has the potential to confuse the execution sequence and make it more
difficult to predict the derived model given a particular input. For
this reason, we feel the benefits afforded by a priority-based execution
model is offset by the added complexity required of the user to effectively
exploit rule priorities.

\subsection{\label{sec:Working-with-Symbols}Boolean Operations on Shapes}

PSML provides unique functions, described in the next paragraph, that
allow programmers to access and operate on derived non-terminal and
terminal shapes from sequential code. This is accomplished using a
\emph{shape path. }The shape path for a specific parse tree node is
created by tracing the ancestry of that node and concatenating the
shape names of each parent in this ancestry using the '/' character
as a name delimiter. For example, in Algorithm \ref{alg:Bricks.psm}
the shape path ``Axiom/wall/Bricks/brick'' refers to the ``brick''
nodes generated as children of the ``wall'' non-terminal. 

PSML provides two access functions that allow sequential code to reference
shapes generated in shape grammars (rule blocks): (1) \emph{$terminals(symbolPath)$
}and (2) \emph{$instances(symbolPath)$. }The parameter $symbolPath$
is interpreted as a ``match any'' regular expression, i.e., the
shapes returned will be all nodes whose symbol path can match, in
order, the elements of the $symbolPath$, e.g., the path ``wall/Bricks''
references all shapes having symbol paths that match the regular expression
``.{*}wall.{*}/.{*}Bricks.{*}'' 

This allows programmers to intuitively reference large and possibly
complex arrangements of shapes using this semantic ``shorthand.''
By default, the search starts in the parse tree at the sub-tree associated
with the invoked method but prepending the string with a '/' character
initiates the search from the parse tree root. Both functions return
an array of shape variables. Context sensitive operations can then
be accomplished using sequential logic that inspects the attributes
of these shapes such as scope, geometry, orientation, appearance,
etc.

\begin{figure}[h]
\noindent \begin{centering}
\includegraphics[height=1.5in]{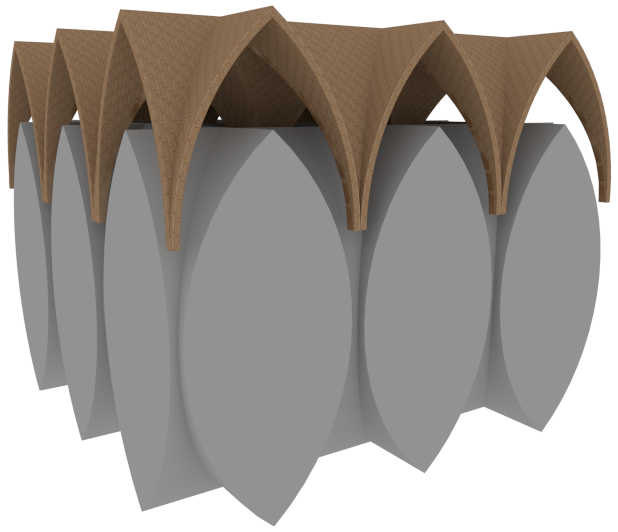}
\par\end{centering}
\caption{\label{fig:Groin-vault-boolean}Boolean operations are used to create
a series of Gothic groin vaults. This is accomplished by creating
a collection of volumes (in gray) that serve to remove bricks from
the overhead ceiling using a Boolean subtraction operation. The remaining
bricks (in brown) form the stones of the groin vault.}
\end{figure}

Two separate access functions are necessary as they provide different
types of references to these nodes. Specifically, the $instances()$
function returns a collection of non-terminal shapes, i.e., shapes
that semantically group together terminals but are not visible in
the final model. The $instances()$ function returns \emph{copies
of }the referenced non-terminal geometries. These shapes can be used
to test to detect context-sensitive situations and to create new shapes.
In contrast, the $terminals()$ function returns \emph{references}
to shapes and allows the user to modify the geometries of all terminal
shapes that are descendants of the referenced shape. At the moment,
shapes returned by the $terminals()$ function are modified by applying
Constructive Solid Geometry (CSG) Boolean operations on the elements
of the associated shape array as (see Fig. \ref{fig:Groin-vault-boolean}).

Since the number and order of symbols generated within a rules block
cannot be tracked or a-priori known in general, both of these functions
\emph{always }return an array of shapes and this array may have length
0. 

Boolean operations may be invoked in sequential code using the function\emph{\footnotesize{}
}\emph{
\[
Shape1.geometricBoolean(Shape2,\,op)
\]
} where\textbf{ $Shape1$ }and $Shape2$ are Shape variables and \textbf{$op$}
is a String that denotes the type of operation. Valid values for $op$
are ``+'' (union), ``-'' (difference), or ``\&\&'' (intersection).
These Boolean operations produce new shapes which may then be queried
by the size of their bounding box using scope variables or by their
volume using the special function volume() which may be invoked on
any shape primitive. Fig. \ref{fig:BooleanDemo.psm-results} shows
how these functions can be used for the context-sensitive placement
of a door based on the presence of a walkway.

Boolean operations typically generate shapes that cannot be modeled
as one of the 19 shapes from Fig. \ref{fig:PSML_PrimitiveShapes}.
Hence, they cannot be used directly as initial symbols for new shape
grammars, i.e., rules blocks. To cope with this issue, rules blocks
that use Boolean shapes as their predecessor symbol are passed the
bounding box of the Boolean shape as the initial shape for the rules
block as seen in Fig. \ref{fig:BooleanDemo.psm-results} when the
bounding box of the intersection is passed to the rule block which
performs the subtraction and replacement with a gate.
\begin{table}
\caption{\label{tab:Effects-of-BooleanOperations}The table summarizes how
Boolean operations impact the geometric shape of symbols as a function
of the type of Shape passed as the 1st and 2nd operand to the geometric
Boolean operation. We refer to the 1st operand as Shape 1 and the
second operand as Shape 2 and NT denotes non-terminal shape variables
(generated by invoking instances()) and T denotes terminal shape variables
(generated by invoking terminals()).}

\centering{}%
\begin{tabular}{|c|c|c|c|}
\hline 
Shape 1 & Shape 2 & Operation & Result\tabularnewline
\hline 
\hline 
NT & NT & any & new NT\tabularnewline
\hline 
NT & T & any & new NT\tabularnewline
\hline 
T & NT & any & T changed\tabularnewline
\hline 
T1 & T2 & +,~\&\& & T1 changed, T2 deleted\tabularnewline
\hline 
T1 & T2 & - & T1 changed\tabularnewline
\hline 
\end{tabular}
\end{table}

The impact of a Boolean operation on a shape variable depends upon
the type of reference associated with both of the involved operands.
When the first operand, i.e., the left-hand operand, is not a terminal
reference, i.e., the first operand is a shape generated by a call
to \emph{instances(), }the Boolean operation generates a new shape
and does not effect geometry of the parse tree shapes\emph{. }When
the first operand is a terminal reference, all of the terminal shapes
referenced by the variable are acted upon with the Boolean which can
change the geometry of the terminal shapes of the parse tree. When
both the first and second operands are terminals and the Boolean operation
is a union or intersection, the geometry of the first operand receives
the result of the Boolean operation and the terminal shapes referenced
by the second operand are deleted. These rules are summarized in table
\ref{tab:Effects-of-BooleanOperations}. Addition of Boolean operations
for modifying terminal shapes and for initiating new rules blocks
enables a rich variety of context-sensitive behaviors and greatly
expands upon the representation capabilities of the language. 

Boolean operations on polygonal models have proven to be difficult
to implement due to numerical instabilities that occur when determining
the intersection of polygonal models. Recent approaches \citep{pavic2010hybrid,zhou2016mesh,barki2015exact}
use exact predicates to determine when and where surface intersections
occur and address many of these shortcomings. In this work, boolean
operations are completed using the Carve software library \citep{CARVE}
which has been used for CSG operations by other open source projects,
e.g., Blender.

\subsection{PSML Development Tools}

\begin{figure*}
\noindent \begin{centering}
\subfloat[{\footnotesize{}(w,h,d)=(15, 20, 36)}]{\noindent \begin{centering}
\includegraphics[height=1in]{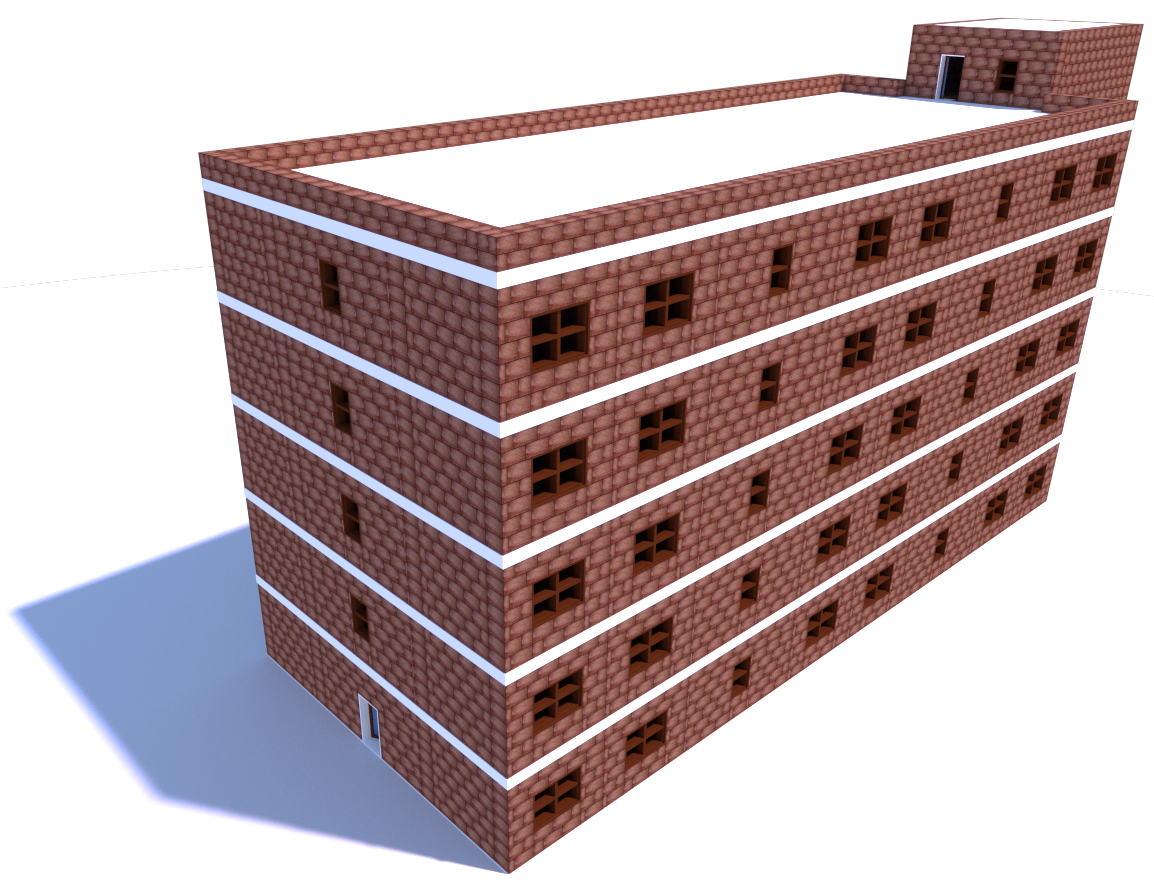}
\par\end{centering}
}\subfloat[{\footnotesize{}(w,h,d)=(20, 10, 10)}]{\noindent \begin{centering}
\includegraphics[height=1in]{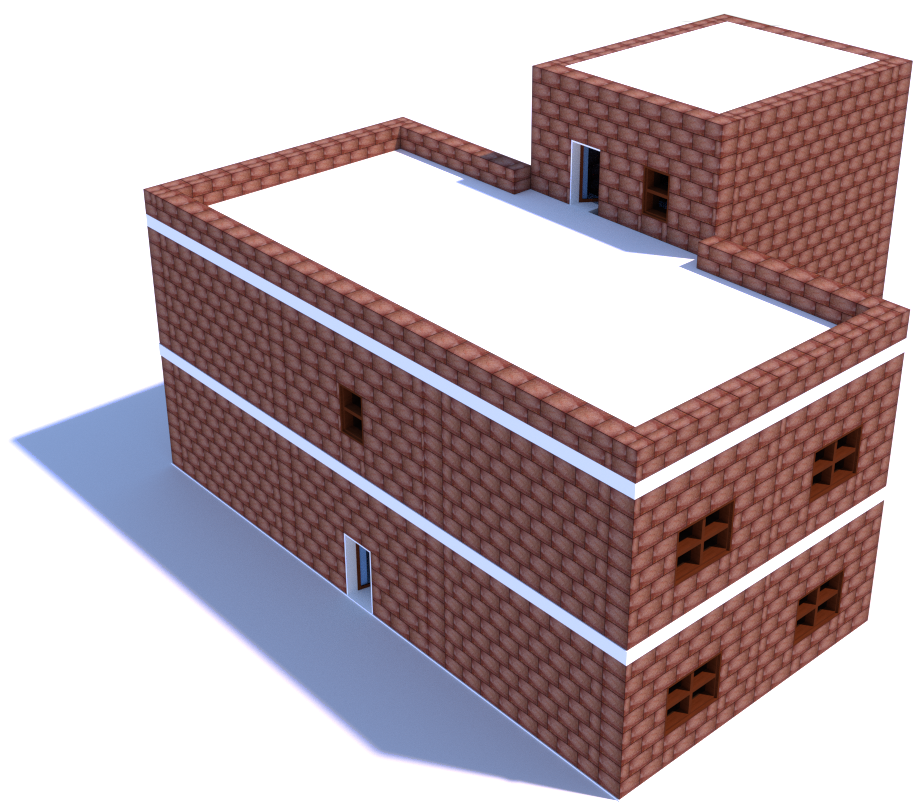}
\par\end{centering}
}\subfloat[{\footnotesize{}The }\emph{\footnotesize{}Stairwell }{\footnotesize{}subtree}]{\noindent \begin{centering}
\includegraphics[height=1in]{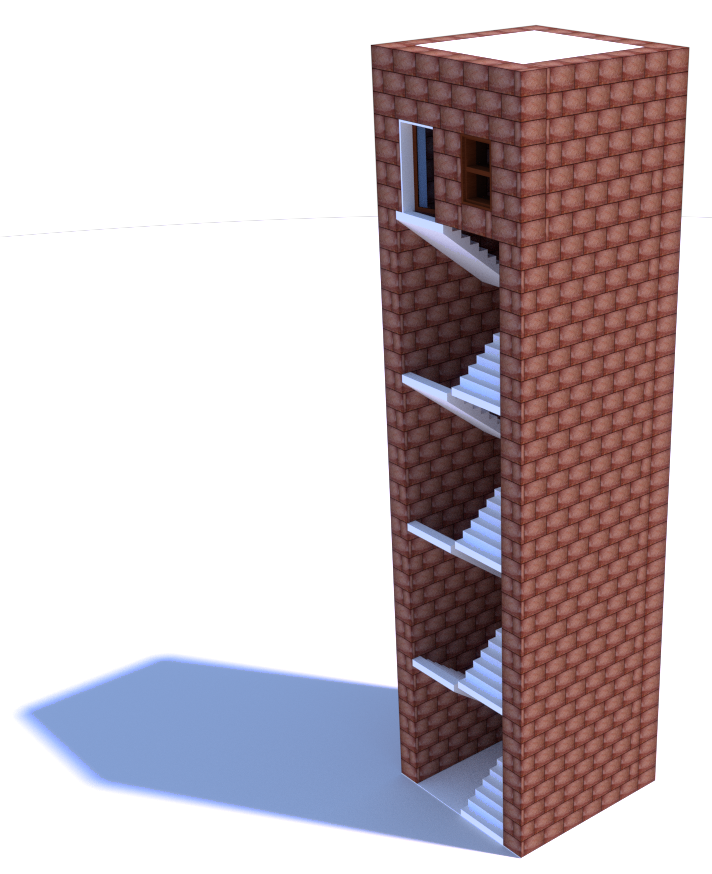}
\par\end{centering}
}\subfloat[{\footnotesize{}An }\emph{\footnotesize{}Office}{\footnotesize{} subtree}]{\noindent \begin{centering}
\includegraphics[width=1in]{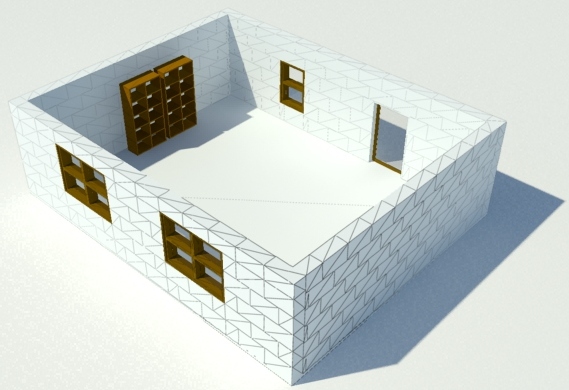}
\par\end{centering}
}\subfloat[\emph{\footnotesize{}Void}{\footnotesize{} spaces colored by function}]{\noindent \begin{centering}
\includegraphics[height=1in]{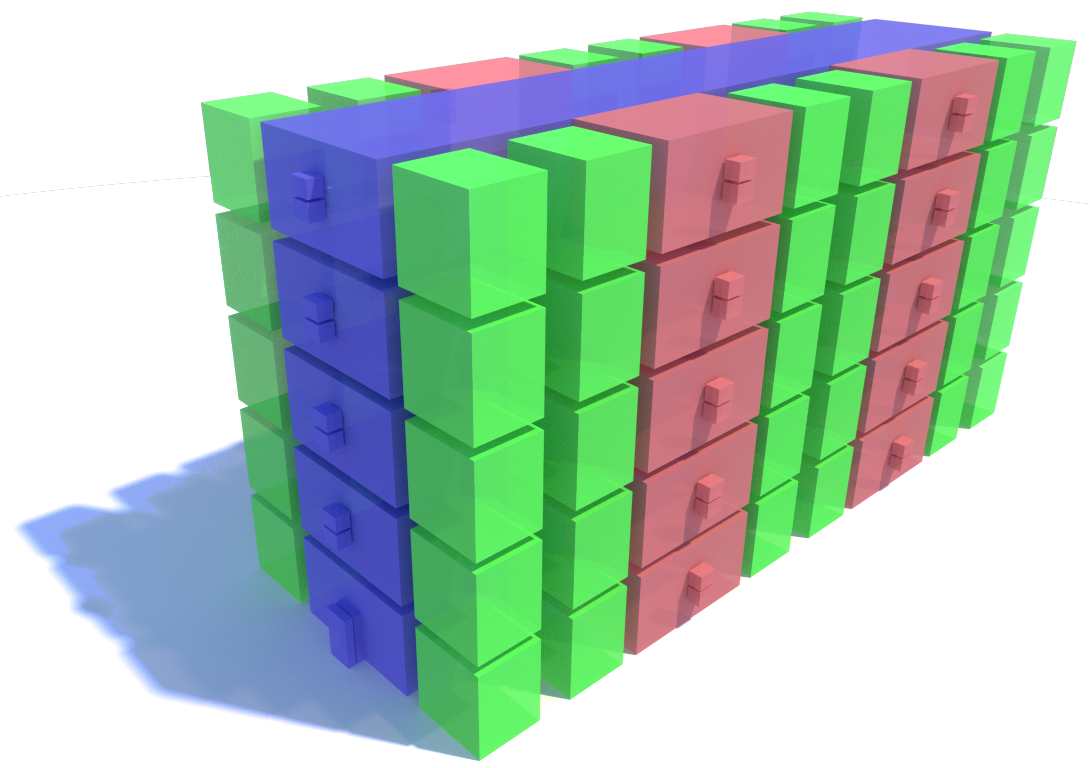}
\par\end{centering}
}
\par\end{centering}
\caption{\label{fig:Office-building}PSML is used to model an office building.
(a, b) show two models generated using different (\textbf{w}idth,\textbf{
h}eight,\textbf{ d}epth) dimensions for the building. (c,d) show sub-structures
extracted as sub-trees of the global parse tree. (e) depicts the 3d
void spaces of the build and color-codes these spaces by function
(see \S~\ref{subsec:Results-Office-Building} for discussion).}
\end{figure*}

We have implemented an Integrated Development Environment (IDE) for
writing PSML code by customizing Oracle's Netbeans software. Through
an ANTLR supported grammar (http://www.antlr.org/), the IDE supports
PSML syntax and programming, which includes syntax highlighting, indentation
and code completion. The prototype Netbeans IDE has an integrated
PSML interpreter and allows users create, edit, run and visualize
models specified using PSML. A collection of PSML programs are also
available for download. These models range from simplistic models
similar to those used for explanations above to more advanced geometries
such as the castle, office building and cathedral shown in Fig. \ref{fig:TeaserFig},
\ref{fig:Office-building},\ref{fig:Cathedral},\ref{fig:VolumetricModels-and-Physics-Engine}.

\section{Results}

Our results are summarized through the detailed analysis of 5 PSML
models: (1) an office building, (2) a medieval castle gate, (3) a
Gothic cathedral, (4) a populated bookcase and (5) a Puuc Mayan house.
All PSML programming and model generation was performed using the
developed Netbeans IDE on a a laboratory desktop workstation with
16GB of RAM and a 2.67Ghz Intel i5 Quad-core CPU running Linux. PSML
models shown in the figures below were exported in Maya's Wavefront
(OBJ) file format and rendered using the open-source 3d modeling software
Blender. 

\subsection{Office Building\label{subsec:Results-Office-Building}}

The PSML code for the office building took approximately 5 hours to
specify and includes 25 distinct dimensions of variation which include
dimensions for doors, windows, stairs, bricks and three alternative
styles for windows, i.e., separate window styles for offices, the
stairwell and the hallways. Fig. \ref{fig:Office-building}(a,b) show
two instances of the PSML office building created by varying the width,
height and depth of the building bounding box. The larger building
(a) took 2.57 seconds to generate (not including the OBJ file exporting
time, same below) and (b) took 1.2 seconds to generate. The PSML code
includes 15 grammar files and 7 of these grammar files are ``generic.''
Generic grammars are shape grammars that can be re-used in other buildings.
Generic grammars used as a part of the office building model serve
to generate the following sub-structures: \emph{Bricks}, \emph{Door}s,
\emph{Window}s, \emph{Frames} \emph{(for Doors and Windows)}, \emph{Stairs
and UStairs} \emph{(two stairways that share a common platform)}.
Fig. \ref{fig:Office-building}(c, d) show renderings of the \emph{Stairwell}
and \emph{Office }non-terminals and depict how large collections of
geometries can be retrieved using relatively simple semantic queries
into the parse tree. Fig. \ref{fig:Office-building}(e) provides a
visualization of the void-space within the office building which consists
of three distinct types: (1) blue regions denote voids associated
with the \emph{East-West hallways}, (2) red regions denote voids associated
with \emph{North-South hallways} and (3) green regions denote \emph{office}
voids.

Use of the\emph{ instances()}, \emph{terminals()} and \emph{volume()}
functions allow us to collect statistics on the frequency and volume
of each semantic element in the models. The statistics for the model
shown in \ref{fig:Office-building}(a) are provided in Table \ref{tab:Office-Building-Table}.
Since these functions are available to sequential code logic, users
can use PSML to construct shape optimization programs which search
the space of shapes described by the PSML program (sometimes called
the language of the grammar) for models that satisfy specific desirable
volume or dimensional criteria.

\begin{table}
\caption{\textbf{\label{tab:Office-Building-Table}}A count of element instances
and volumes for the office building models shown in Fig. \ref{fig:Office-building}(a,b).}

\centering{}%
\begin{tabular}{|c|c|c|c|c|}
\hline 
Name & Num. & Num. & Vol. & Vol.\tabularnewline
\hline 
 & {\footnotesize{}\ref{fig:Office-building}(a)} & {\footnotesize{}\ref{fig:Office-building}(b)} & {\footnotesize{}\ref{fig:Office-building}(a)} & {\footnotesize{}\ref{fig:Office-building}(b)}\tabularnewline
\hline 
\hline 
{\footnotesize{}wall internals} & {\footnotesize{}4} & {\footnotesize{}4} & {\footnotesize{}27.78} & {\footnotesize{}16.02}\tabularnewline
\hline 
{\footnotesize{}stairway steps} & {\footnotesize{}100} & {\footnotesize{}40} & {\footnotesize{}22.00} & {\footnotesize{}10.30}\tabularnewline
\hline 
{\footnotesize{}stairway platforms} & {\footnotesize{}5} & {\footnotesize{}2} & {\footnotesize{}10.15} & {\footnotesize{}4.75}\tabularnewline
\hline 
{\footnotesize{}window, door frames} & {\footnotesize{}2508} & {\footnotesize{}336} & {\footnotesize{}66.69} & {\footnotesize{}8.80}\tabularnewline
\hline 
{\footnotesize{}roof railing parts} & {\footnotesize{}62} & {\footnotesize{}10} & {\footnotesize{}2.60} & {\footnotesize{}0.42}\tabularnewline
\hline 
{\footnotesize{}bricks} & {\footnotesize{}14537} & {\footnotesize{}4078} & {\footnotesize{}2466.3} & {\footnotesize{}721.8}\tabularnewline
\hline 
{\footnotesize{}floors} & {\footnotesize{}7} & {\footnotesize{}4} & {\footnotesize{}1294.86} & {\footnotesize{}250.62}\tabularnewline
\hline 
{\footnotesize{}stairway base} & {\footnotesize{}95} & {\footnotesize{}38} & {\footnotesize{}10.45} & {\footnotesize{}4.89}\tabularnewline
\hline 
{\footnotesize{}ceilings} & {\footnotesize{}1} & {\footnotesize{}1} & {\footnotesize{}22.80} & {\footnotesize{}22.80}\tabularnewline
\hline 
{\footnotesize{}window, door glass} & {\footnotesize{}477} & {\footnotesize{}64} & {\footnotesize{}4.40} & {\footnotesize{}0.68}\tabularnewline
\hline 
{\footnotesize{}exterior molding} & {\footnotesize{}63} & {\footnotesize{}11} & {\footnotesize{}4.52} & {\footnotesize{}0.79}\tabularnewline
\hline 
{\footnotesize{}North-South halls} & {\footnotesize{}100} & {\footnotesize{}0} & {\footnotesize{}1542.82} & {\footnotesize{}0}\tabularnewline
\hline 
{\footnotesize{}East-West halls} & {\footnotesize{}25} & {\footnotesize{}10} & {\footnotesize{}3491.31} & {\footnotesize{}414.83}\tabularnewline
\hline 
{\footnotesize{}office spaces} & {\footnotesize{}60} & {\footnotesize{}8} & {\footnotesize{}1624.90} & {\footnotesize{}645.12}\tabularnewline
\hline 
\end{tabular}
\end{table}

\setlength{\extrarowheight}{0pt}

\subsection{Medieval Castle Gate\label{subsec:Results-Apollonia-Gate}}

The medieval castle gate model is a good demonstration of utilizing
void space to create context-sensitive grammars. This PSML model is
a representation of the specific gate complex at Apollonia-Arsuf.
Apollonia-Arsuf is a fortress constructed approximately 10 miles North
of Tel-Aviv in the 12th century which was razed after being captured
by the Mamluks during the 3rd Crusade in the 13th century. The PSML
code required approximately 15 hours to write and the resulting model
includes 36 dimensions of variation which include dimensions for the
towers, bricks, crenellation, arrow slits, doorways and the gate and
portcullis complex.

\begin{figure}
\begin{centering}
\subfloat[]{\begin{centering}
\includegraphics[height=1in]{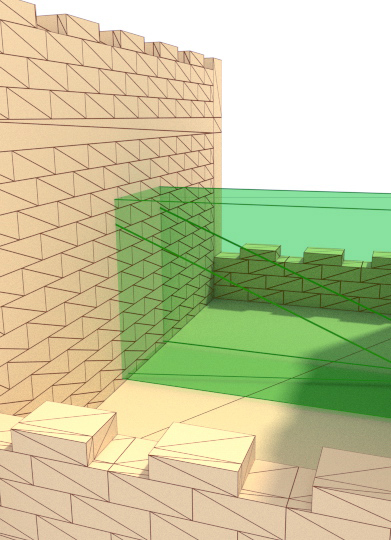}
\par\end{centering}
}\hfill{}\subfloat[]{\begin{centering}
\includegraphics[height=1in]{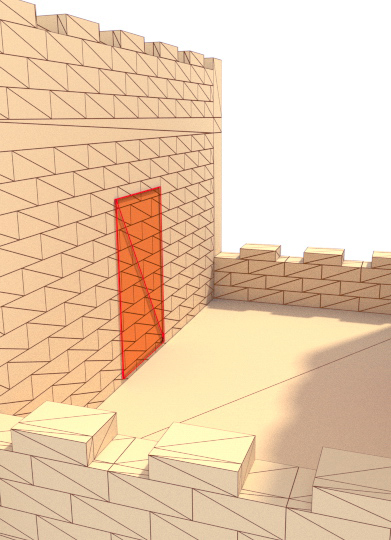}
\par\end{centering}
}\hfill{}\subfloat[]{\begin{centering}
\includegraphics[height=1in]{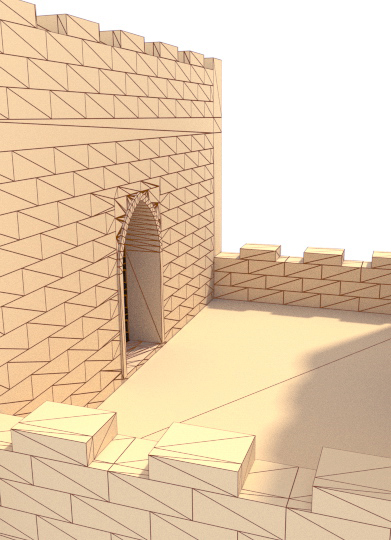}
\par\end{centering}
}
\par\end{centering}
\caption{\label{fig:BooleanDemo.psm-results}(a-c) depicts how void space can
provide context-sensitive geometry generation. (a) shows a part of
the castle terrace from Fig. \ref{fig:TeaserFig}(d) and an adjoining
tower with a walkway void on top of the terrace in green. (c) shows
the Boolean intersection between the walkway void and the adjoining
tower in orange. (d) shows the result of subtracting the intersection
volume and invoking the \emph{GothicDoor()} grammar on the subtracted
volume to replace this volume with a Gothic arch opening.}
\end{figure}

Fig. \ref{fig:TeaserFig}(c,d) show two instances of the PSML castle
gate where the depth of the balcony (overhang) at the top of the tower
and the overall height of the two towers varies. Note that the model
deforms in a meaningful way and parts of it are inserted/removed in
a context-sensitive manner. In particular, when the bridge is at the
same level as the towers, inner crenels are removed as they do not
serve a purpose in positions other than the edges of the roof. In
addition, doors providing access to the bridge are only created when
there is an intersection between the walking space of the bridge and
the tower's side wall. Both of these behaviors are accomplished through
the use of void space. Fig. \ref{fig:BooleanDemo.psm-results} demonstrates
the creation of the doors.

The model shown in (c) took 9.2 seconds to generate and the model
in (d) took 11 seconds to generate. The PSML code includes 12 grammar
files and 5 of these grammar files are ``generic.'' Generic grammars
used as a part of the castle gate model serve to generate the following
sub-structures: \emph{Arch, Bricks, BarrelVault, Crenellation, GothicDoor
}and can be re-used in other Gothic architectural models such as the
Cathedral models in Fig. \ref{fig:Cathedral}. An accounting of the
volumes within the castle gate model is provided in Table \ref{tab:Castle-Gate-Table}.
\begin{table}
\caption{\label{tab:Castle-Gate-Table}A count of element instances and volumes
for the castle gate models shown in Fig. \ref{fig:TeaserFig}(c,d).
Entries with {*} have volumes computed by bounding boxes.}

\centering{}%
\begin{tabular}{|c|c|c|c|c|}
\hline 
Name & Num. & Num. & Vol. & Vol.\tabularnewline
\hline 
 & {\footnotesize{}\ref{fig:TeaserFig}(c)} & {\footnotesize{}\ref{fig:TeaserFig}(d)} & {\footnotesize{}\ref{fig:TeaserFig}(c)} & {\footnotesize{}\ref{fig:TeaserFig}(d)}\tabularnewline
\hline 
\hline 
{\footnotesize{}tower roof} & {\footnotesize{}3} & {\footnotesize{}3} & {\footnotesize{}62.5} & {\footnotesize{}81.3}\tabularnewline
\hline 
{\footnotesize{}bridge roof} & {\footnotesize{}16} & {\footnotesize{}16} & {\footnotesize{}73.75} & {\footnotesize{}46.15}\tabularnewline
\hline 
{\footnotesize{}vault stone} & {\footnotesize{}28} & {\footnotesize{}36} & {\footnotesize{}3.06} & {\footnotesize{}3.94}\tabularnewline
\hline 
{\footnotesize{}arch stone} & {\footnotesize{}8} & {\footnotesize{}12} & {\footnotesize{}2.97} & {\footnotesize{}4.01}\tabularnewline
\hline 
{\footnotesize{}walls} & {\footnotesize{}8} & {\footnotesize{}4} & {\footnotesize{}44.10} & {\footnotesize{}51.98}\tabularnewline
\hline 
{\footnotesize{}bricks} & {\footnotesize{}8870} & {\footnotesize{}12224} & {\footnotesize{}1038.61} & {\footnotesize{}1444.61}\tabularnewline
\hline 
{\footnotesize{}portcullis} & {\footnotesize{}1} & {\footnotesize{}1} & {\footnotesize{}1.86} & {\footnotesize{}1.86}\tabularnewline
\hline 
{\footnotesize{}gate, Gothic door} & {\footnotesize{}4} & {\footnotesize{}6} & {\footnotesize{}61.26 {*}} & {\footnotesize{}73.4 {*}}\tabularnewline
\hline 
\textbf{Totals} & \textbf{\footnotesize{}8938} & \textbf{\footnotesize{}12302} & \textbf{\footnotesize{}1288.11} & \textbf{\footnotesize{}1707.25}\tabularnewline
\hline 
\end{tabular}
\end{table}

\subsection{Gothic Cathedral\label{subsec:Cathedral}}

\begin{figure*}[!t]
\noindent \begin{centering}
\subfloat[]{\noindent \begin{centering}
\includegraphics[height=1in]{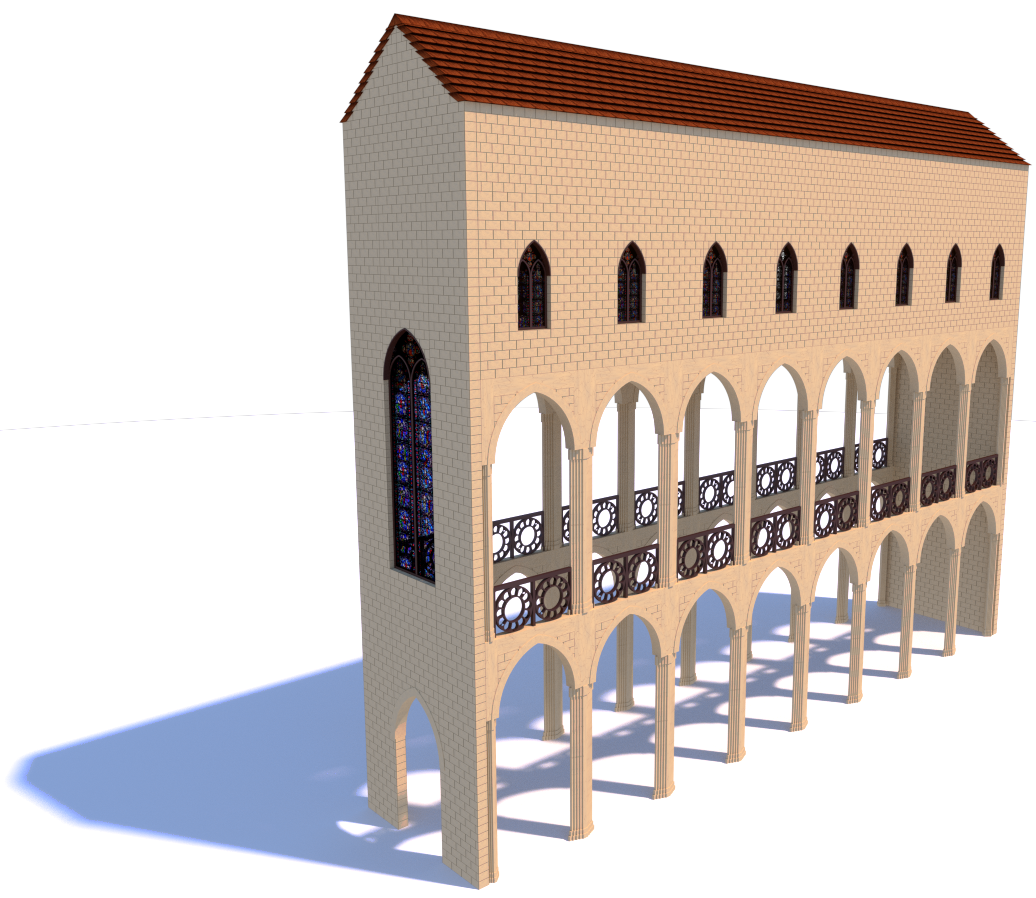}
\par\end{centering}
}\subfloat[]{\noindent \begin{centering}
\includegraphics[height=1in]{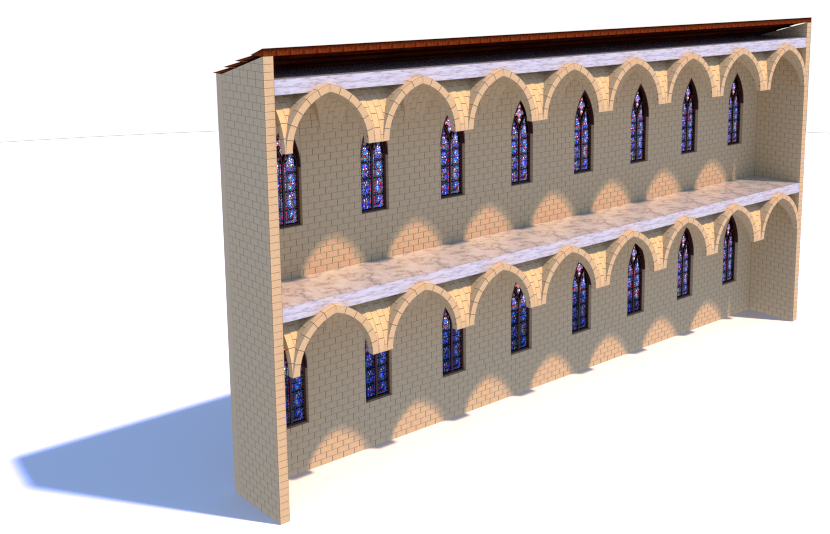}
\par\end{centering}
}\subfloat[]{\noindent \begin{centering}
\includegraphics[height=1in]{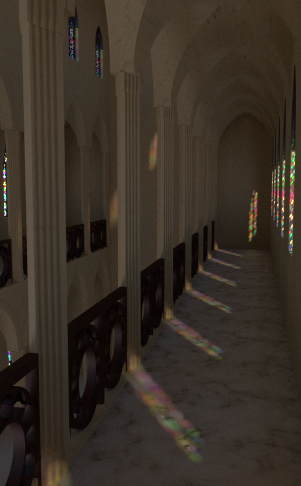}
\par\end{centering}
}\subfloat[]{\noindent \begin{centering}
\includegraphics[height=1in]{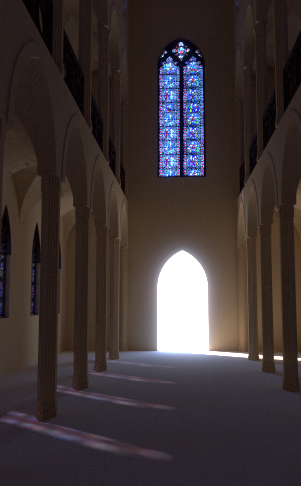}
\par\end{centering}
}\subfloat[]{\noindent \begin{centering}
\includegraphics[height=1in]{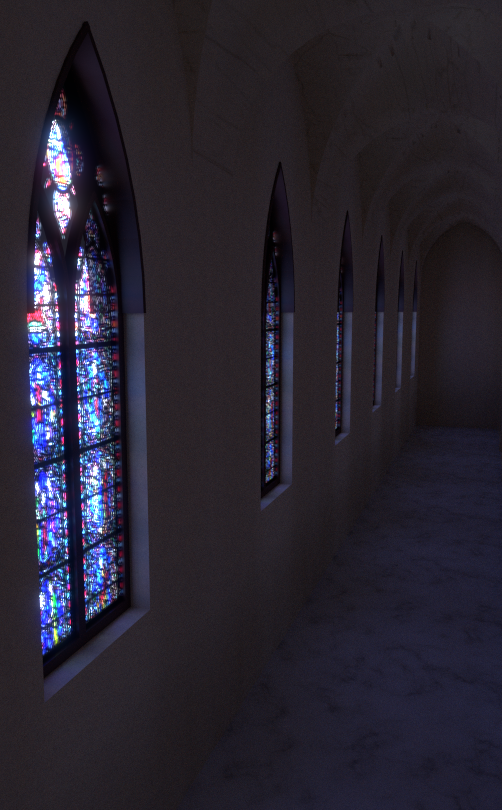}
\par\end{centering}
}\subfloat[]{\noindent \begin{centering}
\includegraphics[height=1in]{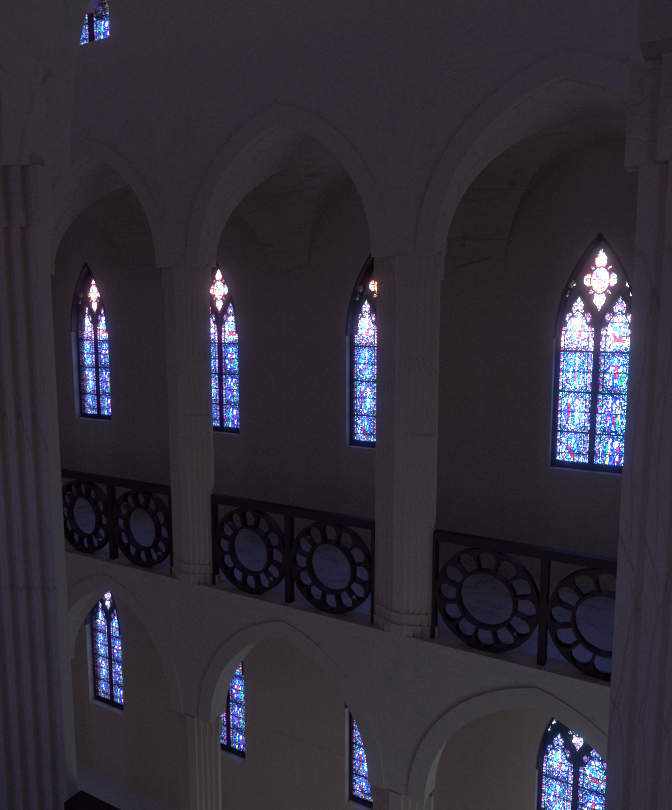}
\par\end{centering}
}
\par\end{centering}
\caption{\label{fig:Cathedral}A PSML model of a Gothic cathedral is shown.
(a) shows an exterior view of a realization of the cathedral. (b,c)
show large-scale components of the cathedral and (d) shows how these
components combine. (e, f, g) show views of the interior of the cathedral
that depict details in the model not visible from the outside.}
\end{figure*}

This PSML code represents the architecture of a Gothic cathedral.
This model, and particularly its interior, is an example of how PSML
can be used to represent complex and highly-detailed architecture.
The code performs many Boolean operations to create the groin vaults
and the Gothic windows, which causes the program to take 1 minutes
and 18 seconds to complete. The model contains 49 Gothic windows (\textasciitilde{}30
Boolean operations per window) and 32 groin vaults (\textasciitilde{}100
Boolean operations per vault). The number of Boolean operations required
is inversely proportional to the size of the vault stones and bricks.
Fig. \ref{fig:Cathedral}(a-f) show sub-tree components of a global
cathedral model (not shown). Images (b,c) show instances of the nave
and aisle of the cathedral as separate components. (d) shows how these
two elements are combined to construct the cathedral model. Fig. \ref{fig:Cathedral}(e,
f, g) are renderings of the interior of the cathedral and highlight
the details present in the interior of the model.

\subsection{Bookcase}

\begin{figure}
\noindent \begin{centering}
\subfloat[]{\begin{centering}
\includegraphics[height=1.2in]{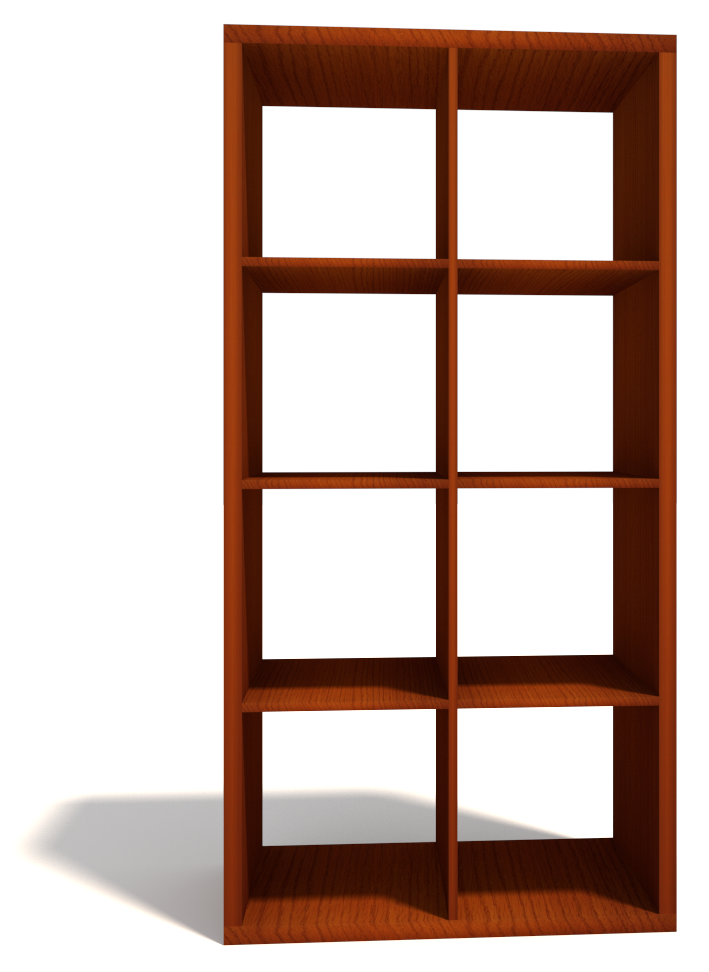}\hfill{}\includegraphics[height=0.75in]{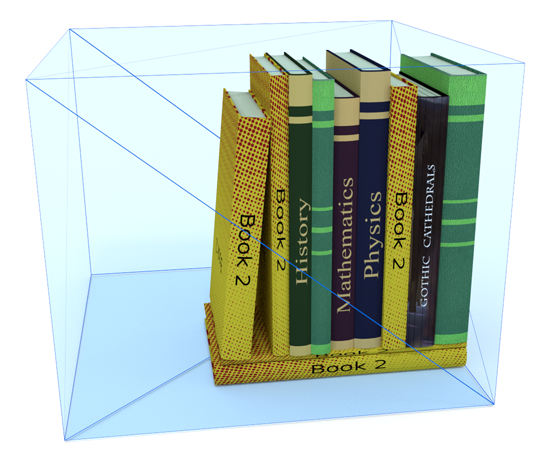}\hfill{}\includegraphics[height=0.75in]{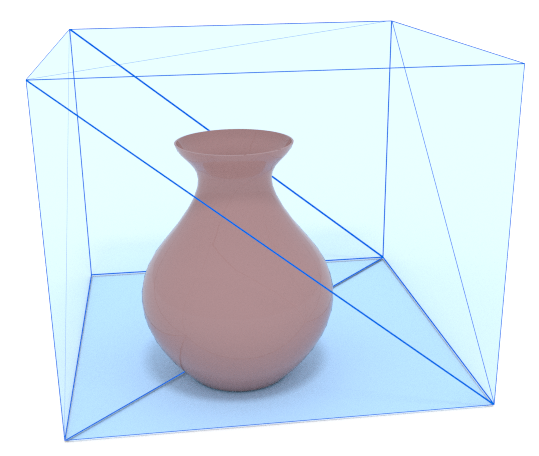}
\par\end{centering}
}
\par\end{centering}
\noindent \begin{centering}
\subfloat[]{\noindent \begin{centering}
\includegraphics[height=1.2in]{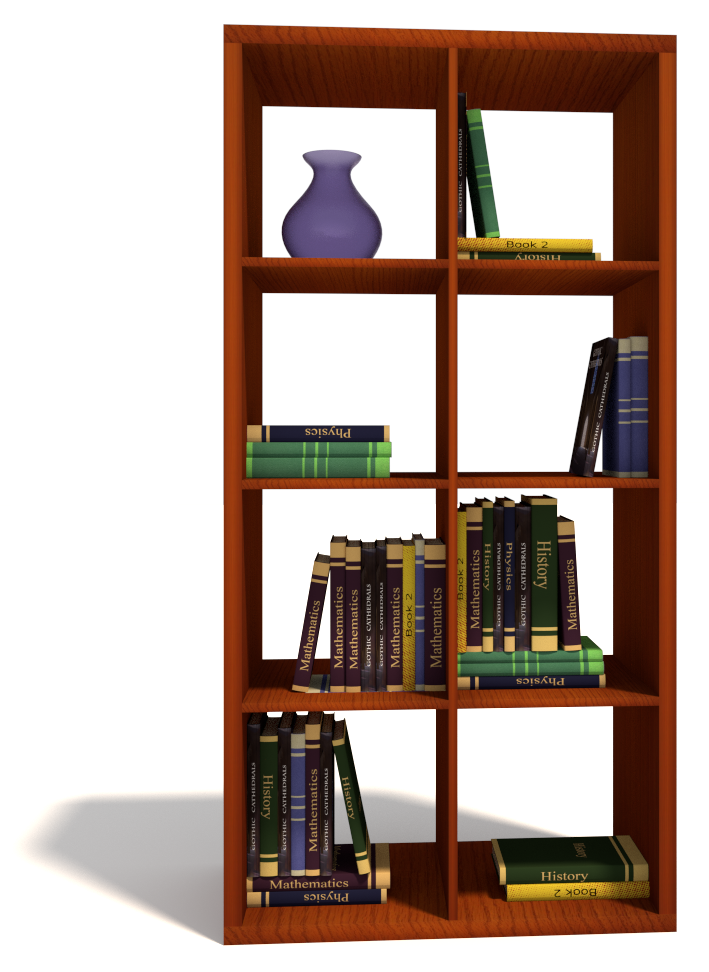}\hfill{}\includegraphics[height=1.2in]{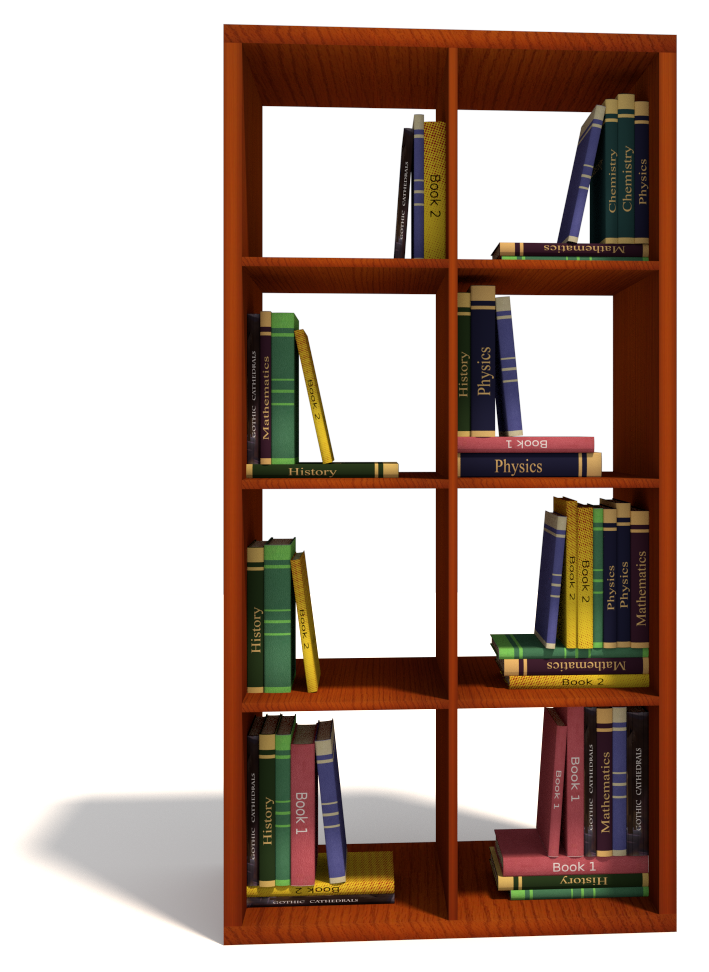}\hfill{}\includegraphics[height=1.2in]{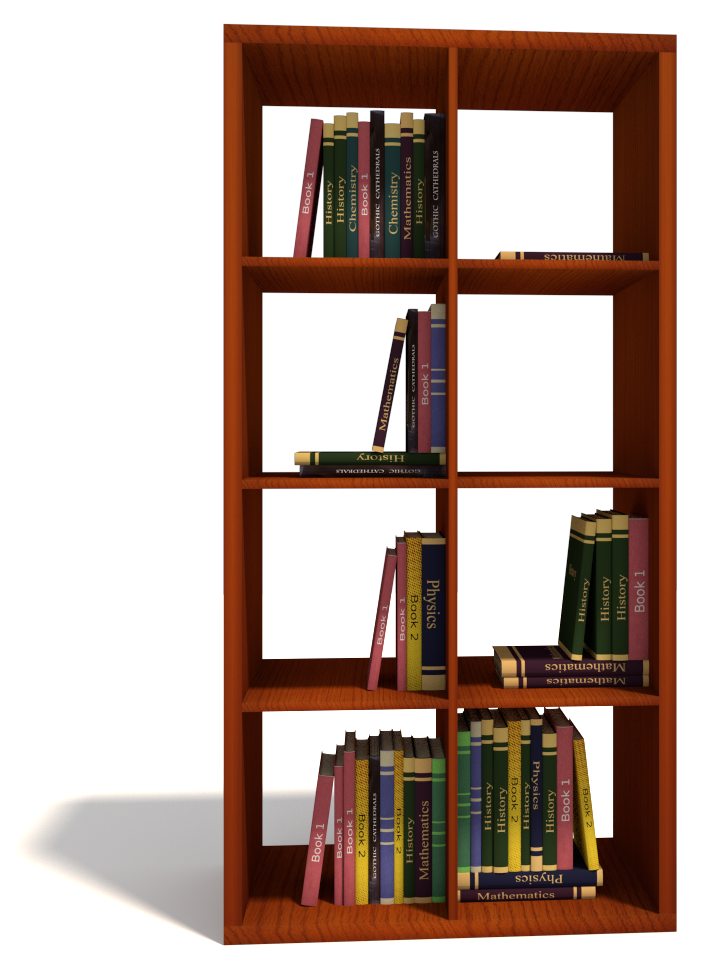}
\par\end{centering}
}
\par\end{centering}
\caption{\label{fig: bookcase}}
 A populated bookcase model generated by combining different independent
components. a) shows each of the components separately. Those components
are, from left to right, an empty shelving unit with labeled space
for the empty cells, a book stack that populates a box (shown in blue)
with a randomized stack of books and a vase, which is randomly placed
inside a box. b) shows three different realizations of the populated
bookcase. The populated bookcase is generated using a fourth grammar
consisting of 5 lines, which creates the empty shelving unit, queries
the cell space, then randomly creates book stacks or vases in some
of the selected cells.
\end{figure}

The bookcase grammar demonstrates the use of object-oriented design
and post-processing to populate empty space. The bookcase model consists
of a shelving unit populated with books and vases, as shown in Fig.
\ref{fig: bookcase}. The shelving unit is generated by a very small
grammar that creates a uniform grid of cells. The books and vases
are generated by two other grammars that populate their scope with
a randomized arrangement of books or vases. In this case, the free-form
vase object is imported as a polygon model (in .OBJ Alias-Wavefront
format) that is substituted into the empty shelf space volume (see
Fig. \ref{fig: bookcase}(a)-right). If the imported model is too
large to fit into the bounding box of their non-terminal parent shape,
it is appropriately scaled-down to meet this constraint. The shelving
unit, books and vases grammars are completely independent and have
been written separately. Another grammar combines them together and
creates the populated bookcase. This shows that PSML is capable of
supporting object-oriented concepts that allow for complicated structures
to be built while keeping individual components relatively simple.

\subsection{Mayan House\label{subsec:Mayan-House}}

\begin{figure*}[t]
\begin{centering}
\hfill{}\includegraphics[height=1in]{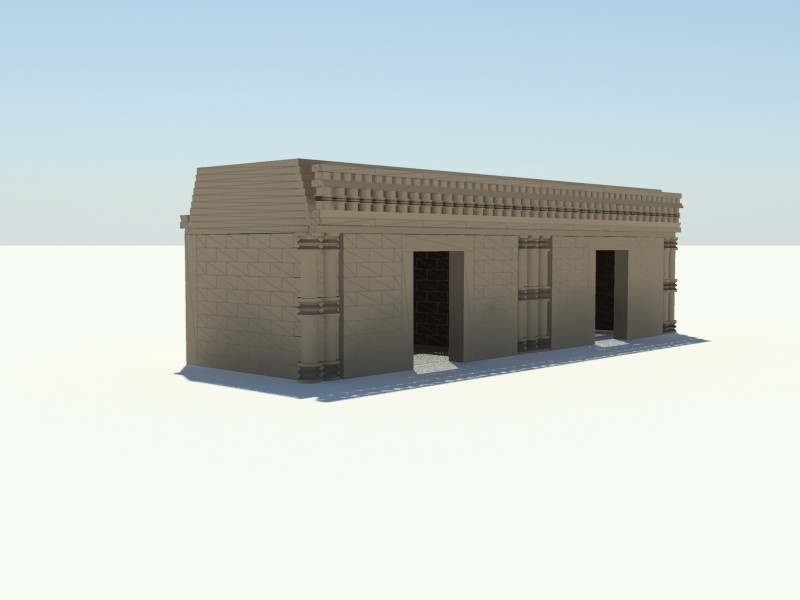}\hfill{}\includegraphics[height=1in]{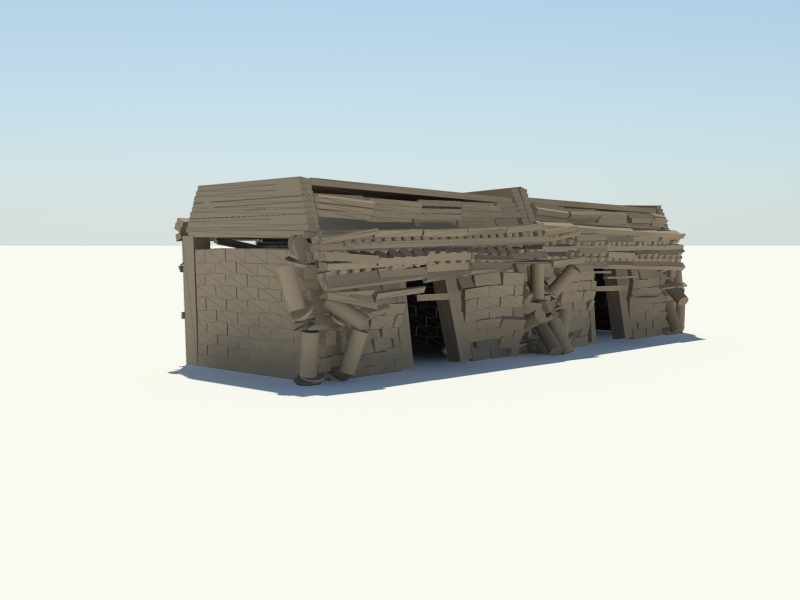}\hfill{}\includegraphics[height=1in]{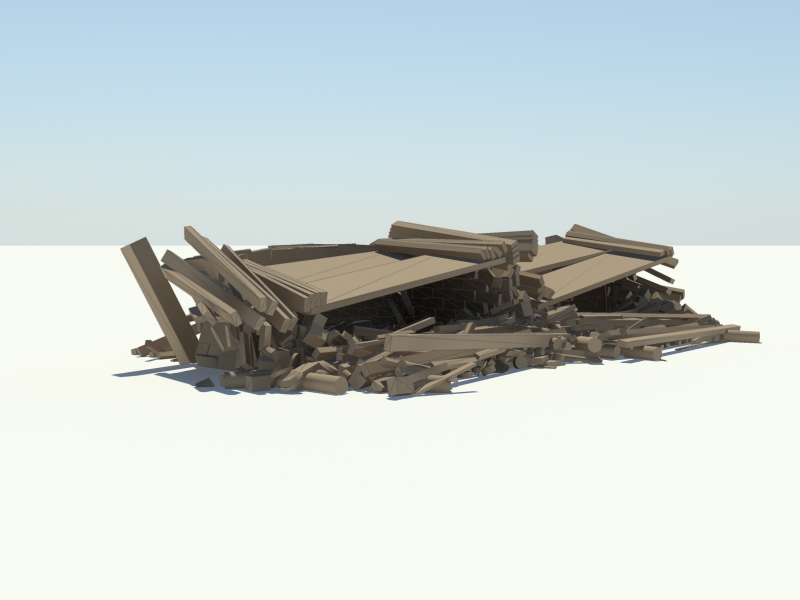}\hfill{}\includegraphics[height=1in]{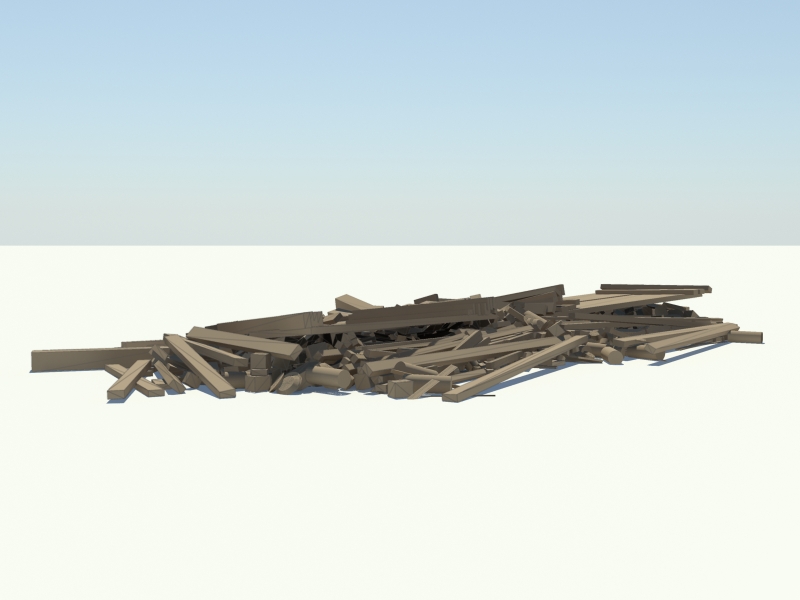}\hfill{}
\par\end{centering}
\caption{\label{fig:VolumetricModels-and-Physics-Engine}Images from left to
right show the physical simulation involving a Mayan building constructed
using PSML programs. Since PSML models consist of 3D volumetric objects,
they can be used directly in physical simulations.}
\end{figure*}

PSML programs represent entire classes of volumetric shape models.
In this example we show how such models can be applied for physical
simulation. Specifically, a PSML model of a Mayan Puuc building from
the classical period is shown in Fig. \ref{fig:VolumetricModels-and-Physics-Engine}(a).
Fig. \ref{fig:VolumetricModels-and-Physics-Engine}(b,c,d) show keyframes
of the model collapse due to a simulated earthquake event (oscillation
of the ground plane). By varying the model derivation, researchers,
e.g., archaeologists, we see possibilities for performing virtual
experiments that may lead to better understandings of how the size/volume
and shape/organization of buildings relate to their collapsed remains
for architectural structures.

\subsection{Programming Effort Analysis}

Statistics of the PSML program code were gathered to approximate the
effort required to generate the models of this article. For our analysis
we tabulate the lines of code required to derive the geometry and
exclude the lines of code written for visualization of the derived
geometry which is also possible using the development environment.
Table \ref{tab:Lines-of-code} includes the lines of code for each
PSML class where the supporting sub-classes are explicitly shown.
PSML sub-classes are group by their semantic purpose (generating walls,
generating roofs, etc.) and each sub-class serves to generate substructures
within the larger model. Their lines of code are listed individually
(column 3) and then summed in the ``total'' column (column 4) of
the table. Subtotals are provided for each structure and give an indication
of the number of code lines required to generate the PSML model. It
is important to note that no effort has been made to optimize the
code and code that is often re-used across structures (e.g., brick
walls) is counted multiple times in Table \ref{tab:Lines-of-code}.
As such, it is entirely possible that the numbers presented could
be significantly reduced using alternative PSML programming approaches.

\setlength{\extrarowheight}{-3pt}

\begin{table}
\caption{\label{tab:Lines-of-code}A summary of PSML code structure and length
for models of this article.}

\centering{}%
\begin{tabular}{|>{\raggedright}p{1.45cm}|>{\centering}p{3.5cm}|>{\centering}p{1.6cm}|>{\centering}p{0.5cm}|}
\hline 
\textbf{\scriptsize{}Fig. Model} & \textbf{\scriptsize{}PSML Functions and Subclasses} & \textbf{\scriptsize{}Lines of Code} & \textbf{\scriptsize{}Totals}\tabularnewline
\hline 
\hline 
{\scriptsize{}Fig. 1(a,b) }\\
{\scriptsize{}\vspace{-2pt}OfficeBuilding } & {\scriptsize{}Building Layout} &  & {\scriptsize{}15}\tabularnewline
\hline 
 & {\scriptsize{}Floor Layout,Cluster, Room, Roof} & {\scriptsize{}(65, 6, 14, 15)} & {\scriptsize{}100}\tabularnewline
\hline 
 & {\scriptsize{}Wall, WallSegment, FullWall, Door} & {\scriptsize{}(28, 27, 19, 38)} & {\scriptsize{}112}\tabularnewline
\hline 
 & {\scriptsize{}Window, WindowFrame, DaylightWindow} & {\scriptsize{}(11, 23, 25)} & {\scriptsize{}59}\tabularnewline
\hline 
 & {\scriptsize{}Stairs, Staircase, StaircasePlatform} & {\scriptsize{}(35, 34, 25)} & {\scriptsize{}94}\tabularnewline
\hline 
 &  & \textbf{\scriptsize{}Total} & \textbf{\scriptsize{}380}\tabularnewline
\hline 
\hline 
{\scriptsize{}Fig. 1(c,d) }\\
{\scriptsize{}\vspace{-2pt}CastleGate } & {\scriptsize{}Building Layout} &  & {\scriptsize{}15}\tabularnewline
\hline 
 & {\scriptsize{}InnerWalls, OuterWall, Corners, Crenels, Doors, Bricks,
Arch} & {\scriptsize{}(56, 54, 25, 78, 13, 10, 26)} & {\scriptsize{}262}\tabularnewline
\hline 
 & {\scriptsize{}Gate, Rectangular Tower, Circular Tower} & {\scriptsize{}(13, 88, 43)} & {\scriptsize{}144}\tabularnewline
\hline 
 &  & \textbf{\scriptsize{}Total} & \textbf{\scriptsize{}421}\tabularnewline
\hline 
\hline 
{\scriptsize{}Fig. 11 }\\
{\scriptsize{}\vspace{-2pt}Cathedral } & {\scriptsize{}Building Layout} &  & {\scriptsize{}355}\tabularnewline
\hline 
 & {\scriptsize{}Barrel vaults, Groin vaults, walls} & {\scriptsize{}(22, 32, 88)} & {\scriptsize{}142}\tabularnewline
\hline 
 & {\scriptsize{}Gothic doors, windows, tracery} & {\scriptsize{}(29, 75, 36)} & {\scriptsize{}140}\tabularnewline
\hline 
 &  & \textbf{\scriptsize{}Total} & \textbf{\scriptsize{}637}\tabularnewline
\hline 
\hline 
{\scriptsize{}Fig. 12 Bookcase} & {\scriptsize{}Layout} & \textbf{\scriptsize{}Total} & \textbf{\scriptsize{}19}\tabularnewline
\hline 
\hline 
{\scriptsize{}Fig. 13 }\\
{\scriptsize{}\vspace{-2pt}MayanHouse } & {\scriptsize{}Building Layout} &  & {\scriptsize{}5}\tabularnewline
\hline 
 & {\scriptsize{}Roof Layout, Column, Walls, Ends} & {\scriptsize{}(122, 61, 19, 26)} & {\scriptsize{}228}\tabularnewline
\hline 
 & {\scriptsize{}House Layout, Doors, Walls, Columns} & {\scriptsize{}(32, 9, 120, 35)} & {\scriptsize{}196}\tabularnewline
\hline 
 &  & \textbf{\scriptsize{}Total} & \textbf{\scriptsize{}429}\tabularnewline
\hline 
\end{tabular}
\end{table}

\setlength{\extrarowheight}{1pt}

\section{Limitations}

The proposed language and methods for generating volumetric models
are limited by the vocabulary of the grammar, i.e., only those shapes
which can be described as a collection of the volumetric elements
from Fig. \ref{fig:PSML_PrimitiveShapes}, can be represented. While
this limits the flexibility of the proposed shape grammar for representation
of completely generic shapes, the breadth of shapes that can be represented
using this grammar is quite large; especially with the incorporation
of CSG Boolean operations on PSML shapes. Current work on PSML investigates
methods to expand the applicability of the proposed language for modeling
geometries that are not well parameterized by a Euclidean, cylindrical,
or spherical coordinate system.

\section{Conclusion}

This article has introduced a volumetric-only approach for procedural
modeling of shapes and a new Procedural Shape Modeling Language (PSML)
that facilitates the implementation of this approach. PSML introduces
a distinct programming construct for procedural model generation that
borrows structure and syntax for sequential statements from Java and
borrows structure and syntax for shape grammars and their production
rules from L-systems. PSML supports an object-oriented approach for
designing models which enables creation of complex models by assembling
simpler models. While creating new models is relatively easy, the
language requires the user to have a basic understanding of object
oriented programming to use the language efficiently. New users can
download the PSML program development IDE, read documentation and
run example code from resources made available at the following GitHub
url: \textbf{\textless{}link removed for review\textgreater{}}. Additionally,
a number of new functions are introduced for use in both shape grammar
contexts and sequential contexts which allow for context-sensitive
generation of highly detailed volumetric models. Such models are of
importance for Building Information Modeling (BIM) systems and the
related issue of closely documenting and tracking the complex structural
state of historic structures. The resulting models are often dense
in terms of their resource utilization, i.e., they often require a
large number of polygons and include details that may not be required
for visualization at a distance. However, this dense collection of
information may be processed by client applications which may extract
portions of the model of interest. The structure of the language and
its execution model is detailed and the application of PSML for generating
context-sensitive volumetric operations and for physical simulation
is demonstrated.

\section*{Acknowledgments}

This research has been partly funded by the US National Science Foundation.

\bibliographystyle{elsarticle-num}
\bibliography{2020_Elsevier_GraphicsAndVisualComputing}

\end{document}